\documentclass[fleqn,10pt,twocolumn, final]{NPGmain}
\usepackage[utf8]{inputenc}
\usepackage[T1]{fontenc}
\usepackage{graphicx}
\usepackage{siunitx}
\usepackage{cancel}
\usepackage{float}
\usepackage{authblk}
\usepackage[switch]{lineno}
\usepackage{lettrine}
\usepackage{soul}
\usepackage[normalem]{ulem}
\usepackage{pdfpages}
\usepackage{orcidlink}
\usepackage{anyfontsize}
\usepackage{siunitx}
\usepackage{xcolor}
\usepackage{physics}
\usepackage{todonotes}
\usepackage{comment}
\usepackage{doi}
\usepackage{wrapfig}

\soulregister\cite7
\soulregister\ref7 

\usepackage{xcolor}

\begin{document}
\title{\LARGE Gigahertz-rate thin-film lithium niobate receiver for time-bin quantum communication
}

\author[1,2,*]{Andrea Bernardi\orcidlink{0009-0007-7994-4651}}
\author[3]{Marco Clementi\orcidlink{0000-0003-4034-4337}}
\author[3]{Marcello Bacchi\orcidlink{0009-0003-8663-6754}}
\author[4]{Matías Rubén Bolaños\orcidlink{0000-0003-3525-0395}}
\author[3,5]{Sara Congia\orcidlink{0000-0003-0287-7669}}
\author[2]{Francesco Garrisi}
\author[2]{Andrea Martellosio}
\author[2]{Marco Passoni}
\author[2]{Alexander Wrobel}
\author[4,6]{Costantino Agnesi\orcidlink{0000-0003-0830-2057}}
\author[4,6]{Giuseppe Vallone\orcidlink{0000-0003-4965-5801}}
\author[4]{Paolo Villoresi\orcidlink{0000-0002-7977-015X}}
\author[2]{Federico Andrea Sabattoli\orcidlink{0000-0002-0029-1755}}
\author[3]{Matteo Galli}
\author[1]{Daniele Bajoni\orcidlink{0000-0001-6506-8485}}

\affil[1]{Dipartimento di Ingegneria Industriale e dell'Informazione, Università di Pavia, Via A. Ferrata 5, 27100 Pavia, Italy}
\affil[2]{Advanced Fiber Resources Milan s.r.l., Via Fellini 4, 20097 San Donato Milanese, Italy}
\affil[3]{Dipartimento di Fisica “A. Volta”, Università di Pavia, Via A. Bassi 6, 27100 Pavia, Italy}
\affil[4]{Dipartimento di Ingegneria dell’Informazione, Università degli Studi di Padova, Via Gradenigo 6B, 35131 Padua, Italy}
\affil[5]{CEA LETI and University Grenoble Alpes, MINATEC Campus, F-38054 Grenoble Cedex, France}
\affil[6]{Quantum Technologies Research Center, Università degli Studi di Padova, Via Gradenigo 6B, 35131 Padua, Italy}
\affil[*]{ Corresponding author. Email: andrea.bernardi01@universitadipavia.it}

\begin{abstract}
Time-bin encoded quantum states of light are crucial for quantum technology applications. 
The integration of manipulation functionalities into chip-scale devices is essential for deploying scalable, high-performance, and cost-effective quantum networks. 
Here we develop a fully integrated, high-throughput quantum receiver based on the thin-film lithium niobate (TFLN) platform, capable of high-speed electro-optic manipulation of time-bin encoded quantum states. 
The device's novel architecture enables active switching of time-bin quantum states with an electro-optic bandwidth exceeding \SI{30}{\giga\hertz}, while supporting real-time arbitrary projective measurements with a bandwidth of over \SI{1}{\giga\hertz}.
We showcase its versatility and performance through several applications, including the certification of entanglement with Bell's inequality violation by 38 standard deviations and with >95\% visibility. 
We then apply it to a fiber-based quantum communication scenario, where we experimentally demonstrate an entanglement-based quantum key distribution (QKD) protocol, achieving stable finite-size secure key rates exceeding 25 kbit/s over 12 hours of continuous operation. 
By leveraging a high-speed active switching scheme, the system overcomes the need for temporal post-selection, eliminating a fundamental loophole that compromises the security of time-bin entanglement-based QKD protocols and relaxes the temporal resolution requirements of single-photon detectors.
Moreover, it enables active selection of the projection basis, increasing the flexibility for communication parties.
This approach establishes a versatile and scalable architecture for time-bin encoded quantum communication, enabling practical protocols on industry-grade photonic technology.
\end{abstract}

\twocolumn
\captionsetup[figure]{labelfont={bf},name={Fig.},labelsep=none}
\maketitle
 \fontsize{10pt}{10pt}\selectfont

\section*{Introduction}
Entanglement is a fundamental physical resource~\cite{Schrdinger1936} that, by the laws of quantum mechanics, enables functionalities that are impossible to achieve with classical systems\cite{Nielsen2012}.
It plays a central role in a wide range of quantum information applications, including quantum communication, computing, and sensing. 
Among these, the development of quantum networks has attracted substantial interest in recent years. 
Such networks facilitate the long-range transfer of quantum information, often leveraging existing telecommunication infrastructure~\cite{Pittaluga2025}, and enable the distribution of cryptographic keys via entanglement-based QKD protocols~\cite{ekert92, bennett92, bennett2014, lo2014secure}, which provide stronger security than traditional “prepare-and-measure” schemes.
Specifically, they allow for secure communication even in the presence of untrusted network components~\cite{bennett92, ekert92} and offer enhanced robustness over long distances~\cite{ma2007quantum}.
In the realm of quantum photonics, entanglement can be implemented using different degrees of freedom of the electromagnetic field, such as path, polarization, frequency, and time --- the latter either in a continuous (“time-energy”)~\cite{Franson1989, tittel1998violation} or discrete (“time-bin”) basis~\cite{Brendel1999, Marcikic2002}.
Each encoding scheme offers specific advantages and trade-offs depending on the application and hardware constraints.
Among them, time-bin entanglement stands out in quantum communication applications for its compatibility with existing fiber-optic infrastructure and resilience to environmental perturbations, making it ideal for long-distance QKD implementations~\cite{montaut2025progress, Singh2025}. 
Time-bin entanglement is relatively easy to generate and manipulate, and supports advanced functionalities such as high-dimensional encoding\cite{yu2025quantum} and hyper-entanglement involving other degrees of freedom\cite{chapman2022hyperentangled, congia2025generation}. 
More generally, time-bin encoding entails a natural way to achieve temporal synchronization among users, automatically establishing a common frame of reference between two or more communication parties\cite{Patel2012}.

Recently, great research interest has been devoted to the development of quantum technological applications based on time-bin entanglement.
On the one hand, efforts have focused on enhancing clock rates --- and more in general secure key rates (SKR) --- in QKD protocols\cite{zhang2008generation, Mueller:24}.
Other studies have investigated the combination of such encoding scheme with dense wavelength division multiplexing (DWDM), aiming at parallelizing multiple quantum channels in star-topology architectures to increase network capacity ~\cite{Fitzke2022, Kim2022, Huang2025}.
Further attention has been devoted to the implementation of time-bin entanglement in integrated photonic devices.
Recently, the programmable generation and the tomography of time-bin-entangled states have been demonstrated on a single thin-film lithium niobate (TFLN) photonic chip\cite{finco2024time, maeder2026programmable}.
The generation of high-dimensional entangled states, or \textit{qudits}, has also been shown through a chip-based interferometers cascade and subsequently exploited to demonstrate high-dimensional entanglement-based QKD\cite{yu2025quantum}.

Despite such great progress, several limitations still hinder the widespread deployment of quantum communication systems based on time-bin entanglement.
First of all, projections on the time-bin basis require high temporal resolution single photon detectors, which should be able to unambiguously discriminate between the “early” and “late” time of arrival of the associated wavepackets, typically limiting the separation of time bins to at least 100s picoseconds.
Even more critically, time-bin entanglement-based applications are affected by the post-selection loophole (PSL)~\cite{Aerts1999}. 
This arises from the necessity to discard non-interfering detection events associated with time-bin wavepackets that do not temporally overlap at the measurement stage. Such discarding compromises the security of QKD implementations~\cite{jogenfors2015hacking, xavier2025energy}, as it allows local hidden variable models to reproduce the observed correlations even in the absence of entanglement by exploiting detector blinding attacks~\cite{Aerts1999}.
Identifying these events requires observers to exchange timing information via classical communication and also necessitates discriminating detection times with a temporal resolution on the order of the time-bin separation.
Thus, overcoming the need for high temporal resolution and closing the PSL are essential objectives, especially in the perspective of using photonic integrated technologies for QKD applications, where low time-bin separation times are sought to minimize the footprint of integrated delay lines and to maximize the SKRs.

Overall, photonic chips would allow scalability, cost-effectiveness, high throughput, reduced size, and high phase stability compared to their bulk optics and fiber-based counterparts, and it would therefore be desirable to foster their deployment in the development of time-bin entanglement based quantum applications. 
To enable this,
it is essential to overcome the limitations imposed by the PSL, improve system integration, and support high-speed operation without increasing demands on detection hardware.

In this work, we address these limitations by developing a fully integrated quantum receiver designed for general time-bin quantum applications.
We resolve the PSL by implementing an ultrafast active switching scheme~\cite{Vedovato2018} exhibiting a \SI{30}{\giga\hertz} electro-optic (EO) modulation bandwidth, made possible by TFLN technology. This allows for the deterministic superposition of all time-bin wavepackets states with picosecond-range separation. Consequently, this eliminates the need for post-selection and, in the case of measurements in the interferometric bases, relaxing the time resolution constraints on the detection system, here ultimately limited only by the clock rate (i.e. by the repetition rate of the pump laser).

The device here presented is a general purpose time-bin universal projector that can be used for a variety of applications. 
It can operate either as a phase-stable state preparation stage, used to tailor a pump pulse shape, or as a receiver, capable of performing real-time arbitrary projective measurements in the time-bin basis, enabled by an EO bandwidth exceeding \SI{1}{\giga\hertz}. 
Fully packaged in a standard telecom casing, it is compact and stable, that makes it suitable for a wide range of quantum applications.
We demonstrate the potential of our device through several key quantum information applications. 
First, we generate a time-bin entangled state by combining one of the realized integrated devices with an integrated silicon waveguide used as a nonlinear source, enabling the generation of arbitrary time-bin quantum states. 
We certify entanglement by violating Bell’s inequality by 38 standard deviations and achieving interference fringe visibility of over 95\% using other two separate quantum receivers.
Second, we perform quantum state tomography, enabled by the capability of our device to implement arbitrary time-bin projective measurements on two independent users.
Finally we implement two versions of an entanglement-based QKD protocol, both in a PSL-free fashion: 
we first show QKD using a “standard” passive basis selection scheme, where fiber beam splitters are used to choose Alice's and Bob's projection bases. 
In these demonstrations, we achieve finite size SKRs exceeding \SI{25.4}{\kilo\bit/\second} --- the highest reported for any time-bin entanglement-based QKD system --- with sustained stability over several hours and with a total channel loss of \SI{0.5}{\decibel}. 
We then fully leverage the potential of the device as quantum receiver, showing how active basis selection can be straightforwardly implemented using a second input port and pseudo-random bit sequence (PRBS) in the perspective of providing full and real-time arbitrariness in the choice of the projection basis.
Crucially, this performance is achieved with picosecond-scale time-bin separation, yet without requiring detectors capable of resolving it. 
By closing the PSL, our system's timing requirements are relaxed entirely to the clock rate of the source laser, paving the way for high-speed, secure, and truly integrated quantum communication.

\section*{Results}

\begin{figure}[]
\centering
\includegraphics[width=1\columnwidth]{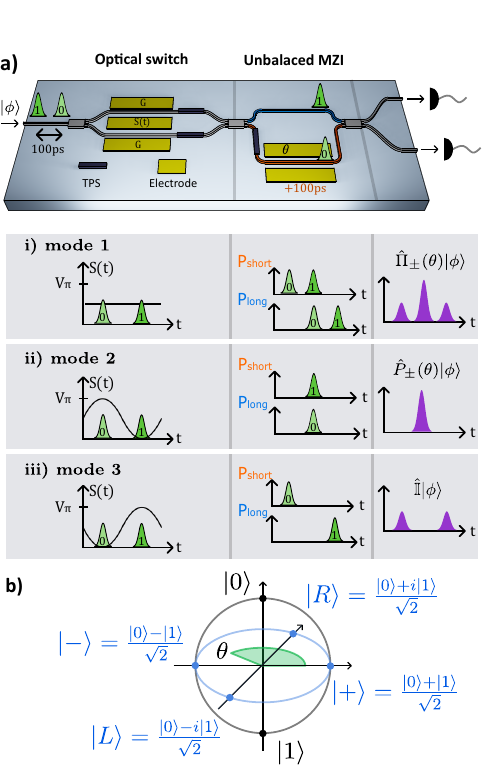}
\caption{
\textbf{ -- Conceptual schematic. } 
\textbf{a} Schematic of the integrated optical circuit. The evolution of the input photon state is illustrated for the three configurations:
(i) the first stage MZM is biased at quadrature with no RF modulation;
(ii) the MZM is driven such that early (late) time-bin photons are routed into the longer (shorter) path of the unbalanced MZI;
(iii) the RF modulation is inverted, routing early (late) photons into the shorter (longer) path.
\textbf{b} The integrated quantum receiver can implement an arbitrary time-bin state projector. When operated in configuration a.ii, it projects into the equator of the Bloch sphere (blue); when operated in a.iii, it projects into the poles of the Bloch sphere.
}
\label{fig1}
\end{figure}

\subsection*{Device description and operating principle}
\begin{figure*}[!t]
\centering
\includegraphics[width=\linewidth]{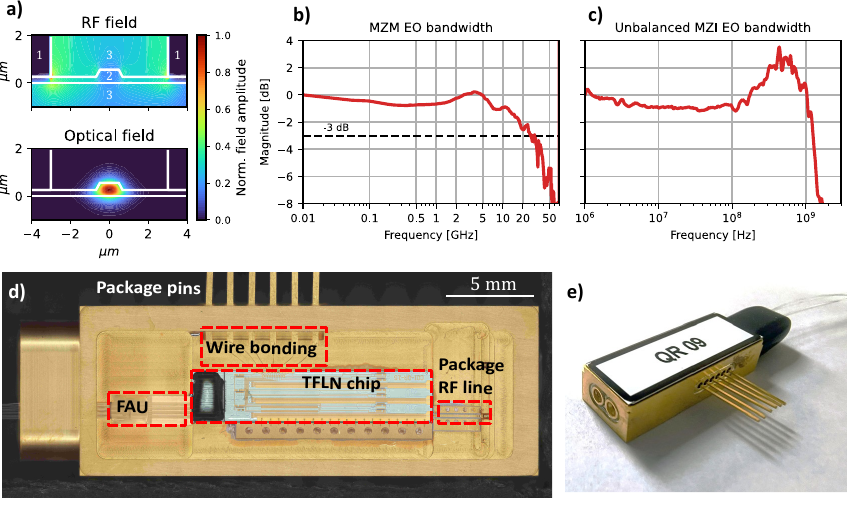}
\caption{
\textbf{ -- Fabricated device. } 
\textbf{a} Simulated transverse component (parallel to the extraordinary optical axis) of the RF and optical electric field in the MZM cross-section. 1: Metal electrodes, 2: Lithium niobate, 3: Silica.
\textbf{b} Measured EO response of the balanced MZM stage and of the \textbf{c} unbalanced MZI stage.
\textbf{d} Image of the realized integrated photonic circuit with \textbf{e} final packaging. FAU: fiber array unit.
}
\label{fig2}
\end{figure*}

A conceptual device layout is shown in Fig.~\ref{fig1}a. 
It consists of two main stages, arranged in a cascaded configuration. 
The first stage is a high-speed balanced Mach-Zehnder modulator (MZM), realized by long and efficient EO phase modulators included in each arm of the interferometer, and exhibiting a broad modulation bandwidth (3-dB bandwidth exceeding \SI{30}{\giga\hertz}; see Fig.~\ref{fig2}b). 
It also includes two thermal phase shifters (TPS), one on each arm of the interferometer, to control the bias operating point while ensuring optical symmetry and extended phase tuning range.
The second stage is an unbalanced Mach-Zehnder interferometer (MZI), where the relative temporal path delay is set to \SI{100}{\pico\second}, obtained by a geometrical path unbalance of \SI{1.31}{\centi\meter}. 
It includes a broadband EO phase modulator and a TPS, both placed in the longer arm of the interferometer, while the shorter arm incorporates a 2\% power-splitter monitor. This component allows for bias monitoring of the MZM and serves to partially compensate for the propagation loss in the longer arm.

The device is designed to operate on a generic qubit encoded in the time-bin basis, schematically depicted on the Bloch sphere in Fig.~\ref{fig1}b, and mathematically described as:
\begin{equation}
\label{eq:qubit}
    \ket{\phi} =
    \alpha\ket{0} +
    \beta 
    e^{i\theta_\mathrm{S}}    
    \ket{1}
\end{equation} 
where $\ket{0}$ corresponds to the photon state in the early time-bin, $\ket{1}$ to photon state is the late time-bin, $\theta_\mathrm{S}$ is the relative phase between the two bins, and $\alpha$, $\beta$ are real-valued probability amplitudes.

As illustrated in Fig.~\ref{fig1}a.i, when the MZM is biased at the quadrature point (\textit{mode~1}), it acts as a 50:50 beam splitter. 
In this configuration, injecting a generic optical field that encodes the time-bin state $\ket{\phi}$ into the device results, heuristically, in a superposition of photons travelling through either the longer or shorter arms of the unbalanced interferometer. 
Photons taking the longer path experience a delay of \SI{100}{\pico\second} relative to those in the shorter path, resulting in a three-peak detection pattern as a function of arrival time after recombination.
Interference arises only in the central peak, where there is the temporal overlap between the early and late time-bin that travel relatively in the longer and shorter paths. 
More rigorously, this configuration implements the positive operator-valued measurement (POVM):
\begin{equation}
    \hat{\Pi}_\pm(\theta) = \tfrac{1}{4}\,\hat{\mathbb{I}} + \tfrac{1}{2}\,\hat{P}_\pm(\theta)
    \label{eq:POVM_mode1}
\end{equation}
where $\hat{\mathbb{I}} = \ket{0}\bra{0} + \ket{1}\bra{1}$ is the identity operator representing non-interfering paths (outer peaks), and:
\begin{equation}
    \hat{P}_\pm(\theta) = \tfrac{1}{2}\,\big(\ket{0} \pm e^{i\theta}\ket{1}\big)\big(\bra{0} \pm e^{-i\theta}\bra{1}\big)
    \label{eq:POVM_mode2}
\end{equation}
is the projector describing the interferometric contribution, associated with the central peak and controlled by the phase $\theta$ applied in the unbalanced MZI.
The signs $\pm$ account for the dependence on the two output ports of the unbalanced MZI stage, which exhibit a relative phase shift of $\pi$.

The device’s innovative aspect arises from its ability to operate the MZM to act as a fast optical switch (\textit{mode~2}): photons in the early (late) time-bin can be deterministically routed into the longer (shorter) arm of the unbalanced interferometer (Fig.~\ref{fig1}a.ii). 
In this way, the input photons belonging to the two time-bin slots are temporally overlapped at the output of the second interferometer. 
This leads to a single interference peak in the detection pattern over time, whereas all photons contribute coherently to the interference pattern.
In these conditions, the implemented POVM is $\hat{P}_\pm(\theta)$, as defined in Eq.~\eqref{eq:POVM_mode2}.
Indeed, by varying the phase $\theta$ in the longer arm of the unbalanced MZI, either via the TPS or the EO modulator, it is possible to project the time-bin state into any equatorial state of the Bloch sphere (Fig.~\ref{fig1}b) without any temporal post selection on the detected events, i.e., discarding contributions from the lateral peaks.
Moreover, the projections can be executed with a high degree of precision from static conditions (tens of hours) with the TPS, and up to GHz speeds with the EO phase modulator. 

A third operating condition (\textit{mode~3}) is possible when the modulation signal applied to the optical switch is phase-shifted by \SI{180}{\degree}, thus reversing the photon routing (Fig.~\ref{fig1}a.iii). 
In this configuration, at the output of the device it is possible to perform a measurement in the computational basis by post-selecting events in the early or late time bins, as now the POVM is $\hat{\mathbb{I}}/2$. 
In this configuration, the time-bin separation at the output doubles from \SI{100}{\pico\second} to \SI{200}{\pico\second} due to the propagation of the late (early) pulse through the longer (shorter) interferometer path. 
The doubled temporal separation relaxes the requirements on the temporal resolution of the detection system, improving compatibility with detectors exhibiting moderate timing jitter.

By comparing the operating conditions of \textit{mode~1} and \textit{mode~2}, we can see that there is a fundamental difference in the achievable interference visibility if no post selection is applied to the detected photons.
When a generic time-bin encoded pure state $\ket{\phi}$ is injected into the device, the maximum interference visibility at the output increases from a maximum of 50\% in \textit{mode~1} to 100\% in \textit{mode~2} thanks to the optical switching configuration.
This distinction becomes even more relevant in the context of time-bin entangled photon pairs. 
Considering a maximally entangled $\Phi$-type Bell state:
\begin{equation}
    \ket{\Phi^+} = \frac{\ket{00} + \ket{11}}{\sqrt{2}}
    \label{eq:phiplus}
\end{equation} 
If each photon of the pair is sent to two separate devices operating in \textit{mode~1} and no temporal post-selection is applied, the resulting two-photon interference visibility is limited to 25\%~\cite{Vedovato2018}.
With this low visibility, the Bell and Clauser–Horne–Shimony–Holt (CHSH) inequalities cannot be violated~\cite{Verstraete2002}, opening a PSL that affects local-realistic tests of quantum mechanics, and thus hindering the security of QKD protocols~\cite{Aerts1999}.
Even in the presence of high-resolution single-photon detectors by performing temporal post-selection, the number of coincidence events contributing to quantum interference is reduced to 1/4 of the total, reducing the SKR potentially achievable.
In contrast, when both devices operate in \textit{mode~2}, the time-bin components are deterministically overlapped, enabling maximum quantum interference. 
This configuration closes the PSL and produces a single interference peak in the time-resolved coincidence histogram, with visibility up to 100\%, while increasing throughput due to the lack of discarded events. 
As a result, it becomes possible to certify entanglement and support secure QKD with a four-fold increase in the SKR. 

\subsection*{Fabrication and classical characterization}
\label{sec_fabrication}

The integrated device was fabricated on a X-cut lithium niobate on insulator (LNOI) wafer.
The single-mode optical waveguide (propagation loss around \SI{0.2}{\decibel\per\centi\meter}) is combined with radio-frequency (RF) traveling-wave coplanar waveguides for EO modulation (Fig.~\ref{fig2}a). 
The optical and RF field profiles are engineered to maximize the modulation efficiency, with a measured $V_\pi L_\pi\approx \SI{3.37}{\volt\centi\meter}$ (at \SI{1}{\mega\hertz}).
Fig.~\ref{fig2}b shows the EO modulation spectrum for the MZM (optimized for high-frequency operation) and Fig.~\ref{fig2}c for the unbalanced MZI stage, with a 3 dB bandwidth as high as \SI{30}{\giga\hertz} and \SI{1}{\giga\hertz}, respectively.
The reduced bandwidth of the second-stage interferometer reflects the fact that this stage was not optimized for high-frequency operation. In particular, compared to the MZM, few differences account for the limited bandwidth performance: it does not include the integrated 50~$\Omega$ termination, leaving the electrodes open, and it is not connectorized with a high-speed RF connector.
The combined effect of the electrodes capacitance and the long wire bonds introduce inductive and capacitive parasitic effects that result in peaking around 500~MHz and a limited bandwidth. 
Comparable performance can be achieved in future iterations by adopting the same design and packaging solution for the MZI as for the MZM.
Further details are provided in the Materials and methods section.

The fabricated photonic chips were packaged for ease of handling, as illustrated in Fig.~\ref{fig2}d and Fig.~\ref{fig2}e, and connectorized with optical, RF, and low frequency interconnections for the respective components.

The final device has a footprint of 9.6 $\times$ 26 mm.
We stress that a key advantage of the TFLN platform lies in the exceptional EO properties of lithium niobate, combined with the high spatial mode confinement of both the optical and electrical fields (see Fig.~\ref{fig2}a), enabled by the integrated photonic design. 
Additional details on the device design, fabrication, and linear characterization are provided in Materials and methods.
For the experimental demonstrations described in the following sections, one or more devices (all exhibiting comparable performances) were combined in complex experimental setups, where they were employed either as quantum receivers, for projection on the time-bin basis, or as pump tailoring stages at the input of an entangled photon source.
Each one was only stabilized in temperature via thermo-electric coolers (TEC), without the need for further active feedback to compensate for interferometers bias phase drift, already minimized through the thermal stabilization and the isolation provided by the package. 

\subsection*{Entanglement certification} \label{sec:certification}
\begin{figure*}[t]
\centering
\includegraphics[width=1\textwidth]{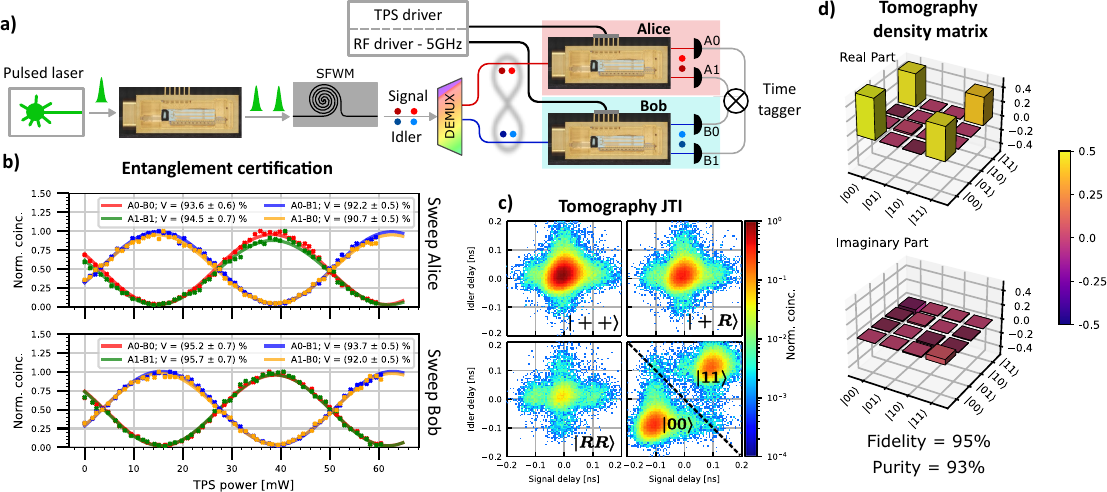}
\caption{
\textbf{ -- Biphoton state analysis. }
\textbf{a} Simplified experimental setup used for data acquisition.
\textbf{b} Measured quantum interference curves without accidental subtraction obtained by varying the unbalanced MZI bias of the first quantum receiver while keeping the second one fixed on the top panel, and vice versa for the bottom panel. 
$V$ denotes the visibility of quantum interference.
\textbf{c} JTI of the output biphoton state for selected projective measurement configurations used for quantum state tomography.
\textbf{d} Experimental density matrix of the generated quantum state computed by maximum likelihood reconstruction algorithm, with projective measurements performed using the TFLN receivers.
}
\label{fig3}
\end{figure*}

We now investigate the suitability of the device, designed as a building block for complex systems involving multipartite states, to manipulate a biphoton time-bin entangled state.
As a first demonstration of its functionality, we certify entanglement by violating both Bell\cite{Franson1989, Brendel1999} and CHSH\cite{clauser1969proposed} inequalities using two spatially separated devices acting as analyzer interferometers while closing the PSL.
The simplified experimental setup is illustrated in Fig.~\ref{fig3}a. 
Optical pulses emitted by an actively mode-locked laser are coupled to one of the fabricated devices, used for pump tailoring.
By tuning the bias operating point of the balanced MZM it is possible to generate a pair of twin pulses separated by \SI{100}{\pico\second}, with arbitrary relative amplitudes. 
These pulses are injected into a \SI{16}{\milli\meter} long integrated silicon waveguide and, under appropriate pump power conditions (see Supplementary Note 4), time-bin entangled photon pairs are generated via spontaneous four wave mixing (SFWM)~\cite{takesue2007entanglement}. 
The resulting quantum state can be expressed as
$\ket{\phi} = \alpha\ket{00}+\beta e^{i\theta_\mathrm{S}}\ket{11}$,
where all the parameters can be freely tuned through the device settings: the probability amplitudes $\alpha$ and $\beta$ by tuning the bias of the first stage MZM, and the relative phase $\theta_\mathrm{S}$ by tuning the phase applied in second stage MZI.
In the following, we set $\alpha=\beta=1/\sqrt{2}$ and $\theta_S=0$, which prepares the Bell state $\ket{\Phi^+}$, defined by Eq.~\eqref{eq:phiplus}.
The output photons are demultiplexed using commercial \SI{100}{\giga\hertz} DWDM filters (ITU channel 28-\SI{1554.94}{\nano\meter} for the signal and channel 38-\SI{1546.92}{\nano\meter} for the idler) and routed to two of the developed devices, hereafter referred to as Alice and Bob.
Each quantum receiver operates in active switching (\textit{mode~2}), as shown in Fig.~\ref{fig1}a.ii, ensuring that all of the input photons contribute to quantum interference.
The two complementary outputs of each interferometer are routed to superconducting nanowire single-photon detectors (SNSPDs), where detection events are recorded using a digital time tagger.

The quantum interference curves retrieved from the coincidence detection patterns for all four detector combinations are shown in Fig.~\ref{fig3}b. 
In the top panel, the phase $\theta_A$ on Alice’s device is scanned while Bob’s phase $\theta_B$ is kept fixed. In the bottom panel, $\theta_B$ is scanned while $\theta_A$ remains constant. 
This measurement effectively implements a Franson-type test of Bell inequality, mathematically corresponding to a projective measurement operated on the state $\ket{\Phi^+}$. 
The expected coincidence rate for outputs at the same side of the second MZI is given by:
\begin{equation}
    \begin{split}
        R \propto {} & \mathrm{Tr} \left[ \left( \hat{P}_{\pm}(\theta_A) \otimes \hat{P}_{\pm}(\theta_B) \right) \ket{\Phi^+}\bra{\Phi^+} \right] \\
        & = \frac{1 + \cos(\theta_A + \theta_B)}{2} 
    \end{split}
\label{eq:bell}
\end{equation}
A $\pi$ shift is applied to the above expression in the case of detections at opposite output ports of the two MZIs.
By fitting the number of accumulated coincidence counts as a function of the applied MZI TPS power with a sinusoidal function, we extract high-visibility two-photon interference fringes, defined as $ V = \frac{\max(R) - \min(R)}{\max(R) + \min(R)} $, which is $\left(93.5 \pm 0.6\right)\%$ on average across all four detector combinations without accidentals subtraction.
For all cases, the fitted visibility exceeds the $1/\sqrt{2}$ threshold required to violate the Bell inequality by at least 38 standard deviations, thus certifying entanglement without any temporal post selection on the detections events.
We also performed the CHSH inequality test by evaluating coincidences at the specific phase settings that maximize the $S$-parameter ($\theta_A = \{ -\pi/4, \pi/4\}$, $\theta_B = \{ 0, \pi/2\}$ for all 4 possible combinations).
We recorded a value of $S = 2.54 \pm 0.04$, exceeding the classical bound ($S=2$) by more than 13 standard deviations.
We stress that the experiment so performed serves at the same time as a test of entanglement, given the violation of the classical bound, \textit{and} of quantum nonlocality, given the use of two distinct devices, albeit not space-like in the present demonstration.
Moreover, both proofs are free of the PSL.
The close match between the experimental data and fits, the similar $\pi$-phase shift powers, and the high fringe visibility across the two devices confirm their stability and reliability.

\subsection*{Quantum state tomography} \label{sec:tomo}
A full tomography of the generated state at the output of the integrated silicon waveguide can be performed by setting the appropriate phase conditions on both integrated quantum receivers simultaneously over a sequence of steps.
A complete quantum state characterization requires projective measurements on the three orthogonal bases for each qubit\cite{altepeter2005photonic}.
In this work, since each quantum receiver can measure the two orthogonal states of a selected basis simultaneously within a single configuration, three measurement configurations were used per device. Permuting the settings between Alice and Bob resulted in a total of 9 combinations, allowing the complete set of 36 projective measurements required for full two-qubit quantum state tomography to be acquired. 
In particular, projections onto the equatorial bases $X=\{\ket{+}, \ket{-}\}$ and $Y=\{\ket{R}, \ket{L}\}$ are performed by operating the MZM to overlap the time bins into a single time slot (\textit{mode~2}, Fig.~\ref{fig1}a.ii) and by adjusting the phase of the unbalanced MZI according to Eq. \eqref{eq:bell}, with $\theta_{A,B}=0$ and $\pi/2$, respectively.
In the computational basis $Z=\{\ket{0}, \ket{1}\}$ the projection is implemented by operating the device optical switch reversely (\textit{mode~3}, Fig.~\ref{fig1}a.iii).

The unitary evolution imparted by the receivers to the biphoton state for such operational modes can be effectively visualized through a measurement of the joint temporal intensity (JTI)\cite{Kuzucu2008, Drago2024, Borghi2024} at the output of the devices.
Fig.~\ref{fig3}c shows the measured JTI of four projective measurements out of a total of nine, for one possible pair of detectors (A0-B0) out of a total of four. 
In the three projective measurements $|++\rangle\langle++|$, $|+R\rangle\langle+R|$, and $|RR\rangle\langle RR|$, we observe a strong suppression of coincidence counts in the outer time bins (i.e., when photons are detected in the earlier or later arrival time slot windows at Alice or Bob), resulting in a well-localized overlap in the central time-bin where interference occurs.
Compared with Eq.~\eqref{eq:bell}, the first and third projectors correspond respectively to the maximum and minimum of the two-photon interference pattern.
In the fourth JTI plot, the projections onto the $\ket{00}$ and $\ket{11}$ states are represented according to which of the two areas, delimited by the dashed lines, the coincidence events belong to. 
Note that, while this measurement requires higher temporal resolution to discriminate the outcome, here the separation of the coincidence peaks is increased by \SI{100}{\pico\second}, leading to a total temporal separation of \SI{200}{\pico\second} and thus relaxing the requirement of detector resolution by a factor of 2.

From the experimental data, we reconstructed the density matrix of the biphoton entangled state through a maximum likelihood estimation algorithm\cite{altepeter2005photonic}.
The result, shown in Fig.~\ref{fig3}d, exhibits a purity of 93\% and a fidelity of 95\% with respect to the Bell state $\ket{\Phi^+}$.
A closer analysis of the density matrix reveals that the generated state deviates slightly from the ideal one. Specifically, the probability amplitude of the $\ket{00}$ component is greater than that of $\ket{11}$, and the imaginary part suggests a slight phase shift relative to the Bell state $\ket{\Phi^+}$.
From the reconstructed density matrix, we provide a further characterization of entanglement beyond the Bell-curve visibility and the $S$-parameter reported in the previous section. 
First, we evaluate the entanglement of formation \cite{wootters1998entanglement} to be as high as 87\%. 
Then, we assess the von Neumann entropy of the reduced subsystems, defined as $E(\rho_{A,B})=-\mathrm{Tr}[\rho_{A,B}\log_2(\rho_{A,B})]$. 
This yields values of $E(\rho_{A})=0.976$ and $E(\rho_{B})=0.978$ for Alice and Bob, respectively, approaching the theoretical maximum of 1, expected for a maximally mixed qubit state. 
These high entropy values indicate a strong degree of mixedness in the reduced states which, when combined with the high purity of the two-qubit state, represent a clear signature of bipartite entanglement \cite{Nielsen2012}.

\subsection*{Quantum key distribution: passive basis selection} \label{2.3}

\begin{figure*}[ht]
\centering
\includegraphics[width=1\textwidth]{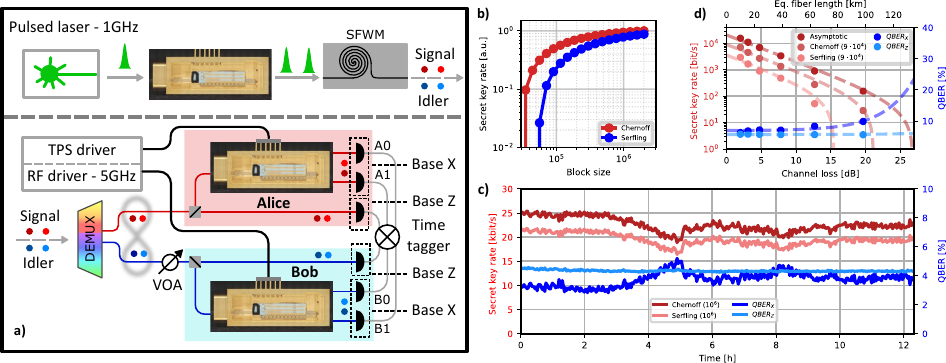}
\caption{
\textbf{ -- Passive basis selection QKD. }
\textbf{a} Simplified experimental setup for QKD with passive basis selection.
\textbf{b} Comparison between the Chernoff and Serfling SKR bounds as a function of the block size.
\textbf{c} In red, the SKR in the finite-key regime ($10^6$ block length) for both the Chernoff and Serfling bounds. In blue, the QBER of the base $X$ and $Z$. 
\textbf{d} Experimental data (dots) and theoretical estimation (lines) of the SKR and QBER for the two measurement bases as a function of the channel link loss. 
The equivalent fiber length is computed considering \SI{0.2}{\dB\per\kilo\meter}.
}
\label{fig4}
\end{figure*}

The integrated TFLN device can serve as a receiver stage for implementing the Bennett-Brassard-Mermin (BBM92) entanglement-based QKD (EBQKD) protocol\cite{bennett92}. 
This is achieved using the same photon-pair source configuration used for entanglement certification and quantum state tomography, with a slight modification introduced at the user stations, as shown in Fig.~\ref{fig4}a.
Specifically, once the photons are demultiplexed and routed towards Alice and Bob, a passive fiber 50:50 beam splitter is placed. 
At one output port, the incoming photons are sent directly to an SNSPD; at the other port, photons are routed to the input of a TFLN device. 
Here, the device is configured with the optical switching activated (\textit{mode 2}), such that the early and late time bins are temporally overlapped at the output (Fig.~\ref{fig1}a.ii). 
The unbalanced MZI is held at a fixed phase $\theta_{A,B} = 0$, enabling projective measurements in the $X$ basis. 
In this configuration, coincidence detection events are maximized between specific detector pairs (e.g., A0–B0 and A1–B1) when the input state is the maximally entangled Bell state $|\Phi^+\rangle$ (Eq. \eqref{eq:phiplus}). 

For photons routed directly to the SNSPDs (bypassing the quantum receiver), a time of arrival analysis distinguishes early ($\ket{00}$) and late ($\ket{11}$) time-bin events, corresponding to projective measurements on the $Z$ basis.
This configuration enables the implementation of the BBM92 protocol, where both users perform measurements in two mutually unbiased bases, $X$ and $Z$. 
The basis choice is intrinsically random due to the passive beam splitter, and the randomness of the measurement outcomes is ensured by the entangled photon-pair source, as demonstrated in the previous section.
For a comprehensive review of quantum cryptography and the various steps involved in QKD protocols, we refer the reader to Ref.~\cite{RevModPhys.74.145}.

In this QKD configuration, there is an asymmetry in detection rates between the $X$ and $Z$ bases due to the additional optical losses caused by the presence of the device in the measurement path of the $X$ basis. The maximum measured coincidence rate is \SI{959}{\hertz} for the $X$-basis and \SI{62}{\kilo\hertz} for the $Z$-basis, with respective optical loss of \SI{13.3}{\decibel} (\SI{15.3}{\decibel}) and \SI{5.9}{\decibel} (\SI{6.3}{\decibel}) for the signal (idler) photons from the entangled photon pairs source to SNSPDs input.
Despite this asymmetry, the QKD protocol can still be successfully implemented by adopting an efficient information reconciliation scheme\cite{lo2005efficient}. 
Under these conditions, the sifting factor $q$ — defined as the probability that both users perform projections in the same basis — is significantly improved, approaching unity. This maximizes the key rate by minimizing the number of discarded events during the sifting process. 
Such an asymmetric basis choice is advantageous, provided that a minimum number of measurements are still performed in the less probable basis (the $X$ basis), to ensure sufficient statistics for detecting potential eavesdropping attempts and thereby guaranteeing the security of the protocol.

To estimate the finite-size SKR, it is required to find an upper bound on the probability of failure on the parameter estimation (PE) step. To do so, we resorted to two different statistical bounds: one based on the Serfling inequality, presented by Tomamichel et al.~\cite{tomamichel2012tight} and from here on referred to as the \textit{Serfling bound}, and a more recent one based on the Chernoff inequality, presented by Mannalath et al.~\cite{mannalath2025sharp} and from here on referred to as the \textit{Chernoff bound} (see Supplementary Note 1). 
Since this work represents the first experimental implementation of the BBM92 protocol using the Chernoff bounds, we opted to show the SKR analysis done with both bounds, to further showcase the advantage of using this new bound with respect to the Serfling one, widely used in literature.
The resulting experimental finite-size SKR as a function of the sifted bit block length is shown in Fig.~\ref{fig4}b. 
Compared to the Serfling bound, the Chernoff bound requires considerably smaller sample sizes for its statistical test to be considered relevant. 
This allows Alice and Bob to generate a secret key at much smaller block lengths, which is especially beneficial for EBQKD, where coincidence rates are usually low.

A long-duration measurement, exceeding 12 hours, was performed running our QKD setup, at $0.5$~dB channel losses. 
The SKR always exceeds \SI{18.8}{\kilo\bit\per\second}, with a maximum value of \SI{25.4}{\kilo\bit\per\second} in finite size regime using the Chernoff bound, with its asymptotic limit value being $\sim14$\% higher, and its finite size regime using the Serfling bound $\sim13.5$\% lower. The quantum bit error rate (QBER) for the two measurement bases and the SKR are reported in Fig.~\ref{fig4}c. 
The QBER in the $Z$ basis is mainly limited by the timing jitter of the SNSPDs, which is $\sim$\SI{50}{\pico\second} full width half maximum (FWHM), compared to a time-bin separation of \SI{100}{\pico\second}.
The detection time window used to distinguish between early and late time bins was optimized to maximize the SKR in the $Z$ basis (see Supplementary Note 2). 
The QBER in the $X$ basis depends on the interferometric performance of the projection implemented by the device. 
Here, an average QBER of 3.76\% was recorded, consistent with the measured visibility reported in the entanglement certification section, according to the relation $\mathrm{QBER}=(1-V)/2$~\cite{RevModPhys.74.145}.
The small variations observed in the QBER in the interferometric basis are mainly attributed to phase drift in the interferometer, which causes deviations of the applied projection from the ideal one.
These fluctuations are reflected in the temporal evolution of the SKR and can be mitigated in practical implementations by standard interferometric stabilization techniques or by periodic protocol interruption and realignment procedures.
Note that other contributions to noise, such as accidental and dark detections, are negligible in the present experimental conditions.

Measurements simulating a variable length fiber link were performed by adjusting the attenuation in Bob's fiber link through a variable optical attenuator (VOA). 
A more realistic proof-of-principle was then performed with fiber spools of different lengths, and could be further extended to a metropolitan-area field test. 
We note that in these scenarios, appropriate dispersion compensation strategies should be put in place to mitigate the effect of chirp for link lengths exceeding \SI{8}{\kilo\meter} (see Supplementary Note 3).
Indeed, group velocity dispersion in standard telecom fibers broadens the time-bin wavepackets, thereby degrading the temporal confinement required for proper device switching and reducing the distinguishability between early and late time bins.
Fig.~\ref{fig4}d shows the measured (dots) and theoretical SKR (dashed line) in the asymptotic regime and for a finite block size of $9\times10^4$.
The corresponding QBER values for the two measurement bases are also shown. 
Note that QKD is feasible for equivalent fiber lengths exceeding \SI{100}{\kilo\meter} (\SI{0.2}{\dB\per\kilo\meter} loss). 
The main limitation in this configuration arises from the signal-to-noise ratio (SNR) of the coincidence counts in the $X$ basis: at high channel attenuation, detector dark counts significantly reduce the visibility of the interference, thus increasing the QBER. 
In contrast, the associated $Z$ basis QBER is almost unaffected by dark counts in the attenuation sweep, due to a higher SNR.
An additional important consideration for real-world implementations involving physically separated stations is the requirement for precise timing synchronization between the users and the pulsed source. 
Such synchronization is crucial to accurately discriminate the received time-bin slots and to correctly align the modulation and switching signals for proper routing within the receivers. 
Efficient synchronization techniques achieving picosecond-level precision have been demonstrated up to \SI{50}{\kilo\meter} link length, even without the need for additional communication channels\cite{pelet2025entanglement, williams2021implementation, calderaro2020fast}.
We note that this synchronization requirement only concerns the technical implementation and does not, in itself, limit or compromise the security of the QKD protocol, whose security relies on entanglement.
However, the security of the present implementation is not device-independent, since the detection loophole remains open \cite{acin2007device}.
Closing this loophole is a well-known experimental challenge, that goes beyond the scope of this work. 
Nevertheless, the proposed overcoming of the temporal PSL represents a necessary step toward the deployment of loophole-free time-bin architectures.

\subsection*{Quantum key distribution: active basis selection} 

\begin{figure*}[ht]
\centering
\includegraphics[width=1\textwidth]{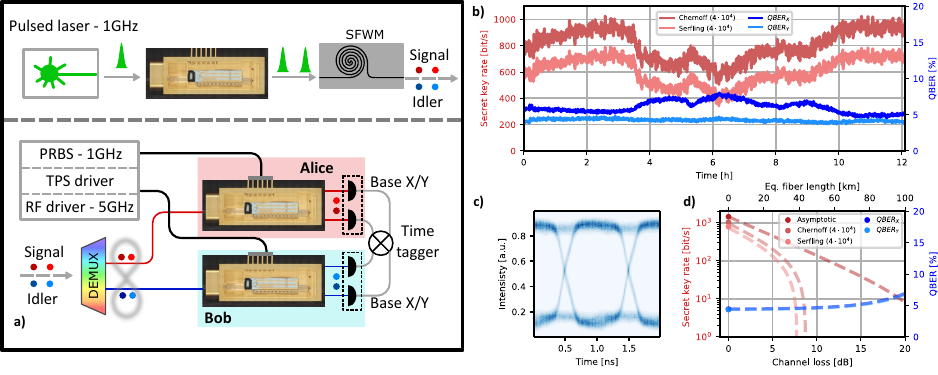}
\caption{
    \textbf{ -- Active basis selection QKD. }
    \textbf{a} Simplified schematic used for the experimental data acquisition.
    \textbf{b} Time trace of the SKR in asymptotic regime and QBER of the tested QKD protocol implementation. 
    \textbf{c} Eye diagram of a single device obtained applying a modulation to the unbalanced MZI with a PRBS at \SI{1}{\giga\hertz} clock rate and peak-to-peak voltage $V_{\pi}$.
    \textbf{d} Theoretical estimation (lines) of the SKR and QBER for the two measurement bases as a function of the channel link loss, based on the experimental acquisition (dots) without added optical fiber stretches.
    The equivalent fiber length is computed considering \SI{0.2}{\dB\per\kilo\meter}.
    }
\label{fig5}
\end{figure*}

The BBM92 QKD protocol can also be implemented using an alternative user receiver setup whose schematic configuration is shown in Fig.~\ref{fig5}a. 
Compared to the previous configuration, which relied on passive basis selection, this configuration removes the beam splitter and directs all the photons toward the device, where measurements are performed by actively selecting the applied projector. 
Precisely, the projective measurements are performed in the two mutually unbiased $X$ and $Y$ bases, both of which exhibit quantum interference for all incoming photons without post selection due to the active time-bin switching (\textit{mode~2}) operated by the first stage MZM. 

The device applies the POVM defined in Eq.~\eqref{eq:POVM_mode2}, corresponding to either $\hat{P}_+(\theta)$ or $\hat{P}_-(\theta)$, depending on the device output port.
The phase $\theta$ determines the measurement basis and is physically implemented by the unbalanced MZI as a combination of two contributions:
\begin{equation}
\theta = \theta_\mathrm{TPS} + \theta_\mathrm{RF}
\end{equation}
where $\theta_\mathrm{TPS}$ is a phase set by the TPS, and $\theta_\mathrm{RF}$ is the phase modulation applied by the EO modulator.

The basis choice for each user is applied at the repetition rate of the pulsed pump laser (\SI{1}{\giga\hertz}) to ensure a random and independent selection for every entangled photon pair.
In our setup the EO modulator of the unbalanced MZI is driven by an arbitrary waveform generator (AWG) programmed with a PRBS.
To improve the security of the protocol, these could, in principle, be replaced by quantum random number generators (QRNG), even integrated on the same chip\cite{peri2026high, bertapelle2025high}.
Alice and Bob use PRBS sequences of lengths $2^7 - 1$ and $2^9 - 1$, respectively, both operating at a \SI{1}{\giga\hertz} bit rate.
This configuration ensures that the joint basis-choice pattern repeats every $127 \times 511 = 64,897$ clock periods, effectively emulating a non-repeating sequence.
The modulation signals have a peak-to-peak voltage of $V_{\text{pp}} = V_{\pi}/2$.
To ensure the switching between desired $X$ and $Y$ measurement bases, the TPS is statically set to apply a phase $\theta_\mathrm{TPS} = \pi/4$ when no voltage is applied to the EO modulator. 
This ensures toggling the total phase $\theta$ between $0$ and $\pi/2$ every clock cycle. An eye diagram showing the continuous-wave light modulation of the unbalanced MZI, driven by the PRBS sequence, is shown in Fig.~\ref{fig5}c.

As for the previous case, we conducted a long-duration measurement exceeding 12 hours continuously without any interruption and without the use of active stabilization, apart from maintaining the device package at a constant temperature.
Once again, both Serfling and Chernoff bounds were used to obtain and compare the secret key rates. 
The time traces of the SKR for both bounds, along with the QBER values for the $X$ and $Y$ bases, are shown in Fig.~\ref{fig5}b.
A maximum SKR of \SI{1.024}{\kilo\bit\per\second} was recorded, with an average of \SI{0.805}{\kilo\bit\per\second} over the measurement period. 
The average QBER value for the $Y$ basis over time is 4.02\%, while the $X$ basis exhibits a higher value of 6.1\%.
For the $Y$ basis, the marginal increase in QBER compared to the previous configuration is primarily attributed to non-ideal AWG modulation, as the basis projection is switched every nanosecond.
The increased QBER observed in the $X$ basis—beyond the limitation affecting the $Y$ basis—is additionally attributed to an initial offset in the operating point, for which the initial QBER was 5.5\%, together with larger phase-drift variations over time.
Nevertheless, given the high modulation speed, the observed QBER values confirm the excellent device performance under fast, active operation.
Compared to the passive QKD system, the SKR is reduced since both measurement bases are now implemented passing through the device and thus the total optical losses are higher: \SI{12.15}{\decibel} for Alice and \SI{10.1}{\decibel} for Bob from the entangled photon pairs source to SNSPDs input. 
Additionally, the sifting factor $q$ is no longer close to 1, but reduced to 0.5, due to the equal probability of selecting each basis.
This high-loss scenario better showcases the difference between the two bounds used to estimate the SKR, as block lengths are smaller.

An estimation of the SKR as a function of channel loss is reported in Fig.~\ref{fig5}d for both the asymptotic regime and finite block size of $4\times10^4$, where using the Chernoff bound provides an increase of $33.2\%$ on the secret key generation when compared with the Serfling bound generation. 
At the measured coincidence rate, a block of key is generated approximately every \SI{25}{\second}. If fast key generation is required, we note that an average of \SI{450}{\bit\per\second} of secret key were generated even using a block length of $10^4$, where the secret key generated using the Chernoff bound becomes $122.3\%$ higher with respect to the Serfling bound. 
At high channel loss, the QBER is primarily limited by detector dark counts. 
In this configuration, compared to the passive one, the QBER degradation for the interferometric bases is expected to occur at slightly longer fiber length, as the receiver coincidence rate increases due to the removal of the beam splitter, which in the previous setup introduced an additional \SI{3}{\dB} loss.

Overall, we emphasize that this configuration, although the SKR is reduced due to the additional optical losses introduced by the quantum receiver and the slight QBER increase arising from the added complexity of the phase modulation, not only removes the PSL as in the passive QKD case, but also eliminates the need to resolve the separation between individual time-bins. 
Indeed, only the separation between successive time-bin states rate must be distinguished.
This relaxed detection requirement makes it possible to increase the SKR by simply raising the pump laser repetition rate and reducing the time-bin separation.

\section*{Discussion}

In this work, we have engineered and demonstrated a fully packaged integrated photonic circuit capable of supporting multiple quantum applications, achieving state-of-the-art performance. 
The main purpose of our device is to overcome the problem of PSL, and the need for temporal post-selection in general, which arises from the nature of Franson-like unbalanced interferometers, and limits both the security and efficiency of QKD protocols based on time-bin entangled states. 
We address this issue by implementing an active switching solution~\cite{Vedovato2018}, of which, to the best of our knowledge, our device represents the first realization on an integrated photonic platform. 
Another integrated solution, proposed by Santangiustina et al.~\cite{Santagiustina2024}, also addresses the PSL issue through a “hug” scheme, that passively discards photons that do not temporally overlap, resulting in a 50\% reduction in usable photon pairs.
More problematically, this approach has compenetrated interferometer that extends along the optical link between Alice and Bob, which hinders the practical deployment of QKD applications in the field.

Our PSL-free architecture can be generally used as an arbitrary time-bin projector and, owing to the combination of TPSs and broadband TFLN modulators, the integrated quantum receiver enables fast switching, in the 10s of gigahertz range, of time-bin encoded states.
This allows the use of closely spaced time-bin states, and therefore high clock (repetition) rates and SKR throughputs, and makes it possible to operate projective measurements at tunable timescales, ranging from sub-nanosecond speeds to stable operation over several hours. 

We have demonstrated its versatility across different tasks: certification of genuine time-bin entanglement by more than 38 standard deviations without the need for temporal post-selection, quantum state tomography of time-bin encoded states, and preparation stage by modulating the pump pulses demonstrating its ability to generate high-fidelity (95\%) and high-purity (93\%) arbitrary time-bin entangled states when combined, as in this case, with a nonclassical light source.
Furthermore, we have also demonstrated that the developed device can effectively implement the BBM92 entanglement-based QKD protocol in both passive and active basis selection modes, proving its versatility and reconfigurability.
The BBM92 is a well established\cite{bennett92, marcikic2004distribution} and widely used QKD protocol based on entanglement. Recent works include high-rate\cite{Honjo:08, Fitzke2022, Mueller:24, Huang2025, pelet2023operational} and high-dimensionality\cite{yu2025quantum} demonstrations. However, the Post-Selection Loophole was always open in these previous works.
To the best of our knowledge, the present work reports the first realizations of time-bin QKD closing the Post-Selection Loophole, achieved at the same time with long-term operational stability (more than 12 hours), and notably, with the highest  SKR (exceeding \SI{25}{\kilo\bit\per\second}) demonstrated so far in time-bin entanglement based systems, as summarized in Tab.~\ref{tab1}.

In the active basis selection QKD configuration, compared to the passive one, the requirement on the minimum time resolution of the detection system is relaxed: now it is sufficient to resolve the time-bin repetition rate, rather than the time-bin separation as in the passive-basis setup.
This was made possible by the unique device's capability to temporally overlap the two time bins through active switching and simultaneously apply a gigahertz-rate modulation in the Franson-like interferometer. 
The fast and accurate control of these modulations enables high visibility of quantum interference, as also confirmed by the low QBER values.
In both QKD demonstrations we report more than 12 hours of continuous QKD operation without any interruption, drift, and the need for active stabilization of the interferometers bias, apart from standard temperature control of the device package following the initial calibration of the required operational phase settings. 

Our results demonstrate that TFLN can be effectively adapted for quantum technologies, enabled by industrial- and commercial-grade fabrication processes.
Indeed, the TFLN platform is an emerging technology originally developed for telecommunications, valued for its high speed EO modulation.
We tailored a complete compact device that integrates multiple high quality optical components originally thought for classical telecom applications -- including edge couplers, beam splitters, EO modulators, thermal phase shifters, and low loss waveguides -- on a single packaged chip, and we combine them to prove their suitability for quantum applications.
Moreover we prove the technology reliability by repeatable excellent performances even across different fabricated devices.
We also demonstrate the first experimental implementation of the BBM92 protocol, taking advantage of the recently found Chernoff\cite{mannalath2025sharp} bound for SKR estimation, proving its advantage with respect to the previously-used Serfling\cite{tomamichel2012tight} bound when small block lengths are used, as is commonly the case for EBQKD protocols.

The present integrated optical circuit can also be extended to other quantum applications.
For instance, recently developed integrated interferometers for multidimensional time-bin states (up to 8 levels)~\cite{yu2025quantum} could be combined with our switching approach to overcome the PSL. 
This would require multi-port interferometers with path delays in multiples of the time-bin spacing and multi-port switching via cascaded high-speed modulators.
The realized device can also be directly employed for the implementation of entanglement swapping~\cite{de2005long}.
Furthermore, it can be combined with the processing of other degrees of freedom, such as frequency and polarization, by exploiting relatively fast EO modulators~\cite{tagliavacche2025frequency} and polarization filters~\cite{zhang2024compact}. This integration could enable hyperentangled quantum applications, including quantum secure direct communication (QSDC)\cite{sheng2022one, yang2025300}, entanglement purification\cite{sheng2010deterministic} and improve channel capacity. 

Perspective device improvements primarily entail increasing the SKR. 
This could be firstly pursued by operating the protocol at higher clock rates. In the case of the active basis selection configuration, it is currently limited to \SI{1}{\giga\hertz} for practical reasons (the limited bandwidth of the unbalanced MZI modulation stage, not optimized for broadband operation, as discussed in the Materials and methods section), but potentially compatible with up to \SI{5}{\giga\hertz}, the maximum repetition rate set by the MZI unbalance delay.
Conversely, in the case of passive basis selection configuration for the investigated channel losses in this work–particularly in the low loss regime–the achievable SKR is limited by the maximum detection rate of the SNSPDs ($\sim$\SI{1.5}{\mega\hertz}). If this limitation is overcome, the repetition rate could also reach \SI{5}{\giga\hertz} without further hardware constraints.
We stress that in our approach, when combined with state-of-the-art low-jitter SNSPDs\cite{Korzh2020} and a reduced unbalanced interferometer delay, the maximum clock rate achievable is ultimately limited by the pulse duration, which in turn determines the minimum time-bin separation required to preserve orthogonality of the states.
In particular, by employing broader-bandwidth optical pulses with picosecond-scale duration, this approach could potentially reach clock rates up to a hundred of gigahertz, and a proportional increase in the SKR.
A second aspect affecting the SKR, particularly severe in EBQKD protocols, are the overall setup losses. 
The optimization of the whole setup, including the design of the fabricated devices, will allow to a direct increase of the coincidence rates, and hence of the overall SKR.

To improve the security of the protocol in the active basis selection configuration and in real application scenarios, the PRBS modulation could be substituted by a QRNG\cite{bertapelle2025high}, potentially realized through compact hybrid-integrated circuits\cite{peri2026high}, in order to guarantee certified randomness of the basis choice.

Further improvements concern the visibility of the interference fringes, where we achieve approximately 93.5\% on average.
This causes an increase in QBER, which ultimately limits the maximum range achievable for QKD.
The primary limitation is attributed to polarization mode mixing, namely the coupling between TE and TM modes in the devices. 
This can be mitigated by integrating a polarization filter\cite{zhang2024compact}, which we have found to improve the extinction ratio from a tested median value of 19.3 dB to 34.2 dB in interferometric performance (see Supplementary Note 5); and using Euler bends~\cite{jiang2018low} in future design improvements.
Another suitable pathway may consists in the use of materials characterized by less severe birefringence, such as the emerging thin-film Lithium Tantalate (LiTaO$_3$) platform, exhibiting comparable Pockels coefficient \cite{wang2024lithium}.
Other minor limiting factors for the reduced visibility are the CAR value, which is around $\sim 100$ for the presented experimental demonstrations, the residual unbalanced losses in the MZI stage, that can be mitigated with the adoption of variable attenuators \cite{finco2024time}, the non-ideal sinusoidal photon-switching modulation, and uncertainties in the applied bias and $V_{\mathrm{pp}}$ modulation signal.

In the perspective of scaling to large and multi-party optical networks, all the experimental implementations here reported were carried out using the standard ITU \SI{100}{\giga\hertz} bandwidth allocation for each of the two user channels, which also matches the transform-limited photon bandwidth and optimizes spectral density. The quantum receiver platform also relies on components designed for C+L band operation, ensuring compatibility with DWDM systems and paving the way for scalable quantum networking over existing telecom infrastructure~\cite{Kim2022, Fitzke2022, Mueller:24, Huang2025}.
By adopting quantum information density as a figure of merit, proposed by Ref.\cite{yu2025quantum}, define as $ d^N / (\Delta t \cdot \Delta v)$ our implementation achieves a value of $2^2/[\SI{100}{\pico\second}\cdot2 \cdot(\SI{100}{\giga\hertz}\cdot 2)]=0.1$. 
This value is among the best results reported in the literature, underlining the efficiency of our integrated solution in terms of both spectral and temporal resource utilization.

In summary, our work demonstrates a versatile and high-performance integrated photonic circuit based of TFLN platform that allows the tailoring of pump pulse for the generation of time-bin states and overcoming the PSL limitation, certifies genuine time-bin entanglement, and enables stable, high-rate EBQKD, paving the way toward scalable quantum communication networks.

\begin{table*}[ht]
\centering
\begin{tabular}{@{}lcccccc@{}}
\toprule
User implementation & DoF & SKR per channel & Link length & Clock rate & Bin separation & Protocol \\ \midrule

Silica-based\cite{Honjo:08} & Time & \SI{0.14}{\bit\per\second} & \SI{100}{\kilo\meter} & \SI{0.333}{\giga\hertz} & \SI{1}{\nano\second} & BBM92\cite{bennett92} \\ \midrule

Fiber-based\cite{Fitzke2022} & Time 
& \begin{tabular}[c]{@{}l@{}} \SI{42}{\bit\per\second} \\ \SI{29}{\bit\per\second} \end{tabular} 
& \begin{tabular}[c]{@{}l@{}} \SI{47}{\kilo\meter} \\ \SI{60.5}{\kilo\meter} \end{tabular} 
& \SI{220}{\mega\hertz} & \SI{3}{\nano\second} & BBM92\cite{bennett92} \\ \midrule

Fiber-based\cite{Kim2022} & Time & \SI{5.1}{\bit\per\second} & \SI{0}{\kilo\meter} & \SI{18}{\mega\hertz} & \SI{3.6}{\nano\second} & E91\cite{ekert1991quantum} \\ \midrule

PIC\cite{yu2025quantum} & Time 
& \begin{tabular}[c]{@{}l@{}} \SI{2.04}{\kilo\bit\per\second} \\ \SI{37}{\bit\per\second} \end{tabular} 
& \begin{tabular}[c]{@{}l@{}} \SI{0}{\kilo\meter} \\ \SI{60}{\kilo\meter} \end{tabular} 
& \SI{250}{\mega\hertz} & \SI{64}{\pico\second} & BBM92\cite{bennett92}\\ \midrule

Free space\cite{Huang2025} & Time & \SI{5.1}{\kilo\bit\per\second} & \SI{0}{\kilo\meter} & \SI{80}{\mega\hertz} & \SI{1}{\nano\second} & BBM92\cite{bennett92} \\ \midrule

Fiber-based\cite{zhuang2025ultrabright} & Pol. & \SI{223}{\bit\per\second} & \SI{201}{\kilo\meter} & CW & --- & modified BBM92\cite{lim2021security} \\ \midrule

Fiber-based\cite{zhong2024hyperentanglement} & Time + Pol. & \SI{700}{\bit\per\second}* & \SI{0}{\kilo\meter} & \SI{50}{\mega\hertz} & \SI{5}{\nano\second} & custom\cite{doda2021quantum} \\ \midrule

Fiber-based\cite{pelet2023operational} & Time-Energy & \SI{7}{\kilo\bit\per\second} & \SI{50}{\kilo\meter} & CW & \SI{1.6}{\nano\second} & modified BBM92\cite{neumann2022continuous} \\ \midrule

\textbf{TFLN (this work)} & \textbf{Time} & \textbf{25.4 kbit s$^{-1}$} & \textbf{0 km} & \textbf{1 GHz} & \textbf{100 ps} & \textbf{BBM92}\cite{bennett92}\\

\bottomrule
\end{tabular}
\caption{
Comparison among entanglement-based QKD demonstrations using different degrees of freedom (DoF). 
All implementations employ fiber-based channels, with channel losses scaling approximately as 
$\sim$ 0.2 dB/km attenuation in standard single-mode fiber for the reported distances.
PIC: Photonic Integrated Circuit, *: Values extracted from published plots.
}
\label{tab1}
\end{table*}

\section*{Materials and methods}
\noindent {\bf Device fabrication and characterization.} 
The integrated photonic devices were fabricated by Advanced Fiber Resources Milan s.r.l. on a 4-inch X-cut LNOI wafer.
Optical waveguides featuring \textit{rib} geometry are patterned by a deep UV (\SI{248}{\nano\meter}) stepper lithography (ASML PAS 5500/350C), on a \SI{550}{\nano\meter}-thick LN layer with a \SI{275}{\nano\meter}-deep etching and are embedded in a silica cladding, supporting single-mode operation across the standard C+L telecom bands, with typical propagation losses of \SI{0.2}{\dB\per\centi\meter}.
The cladding thickness ensures full confinement of the optical mode within the LN layer and the surrounding glass matrix.

A schematic of the fabricated chip is shown in Fig.~\ref{fig1}a. Input photons are coupled into the circuit via one of four edge couplers, and initially pass through a 50:50 multimode interferometer (MMI) that forms the first stage MZM, which functions as an optical switch.
The MZM comprises two geometrically balanced optical arms, each equipped with a TPS for precise bias control. EO modulation is applied through \SI{1}{\centi\meter}-long ground-signal-ground RF electrodes that drive the waveguides in push-pull configuration, enabling high-speed modulation.

The second stage of the device consists of a 2×2 MMI that connects to an unbalanced interferometer with a fixed delay of \SI{100}{\pico\second} between its two arms. The phase in the longer arm can be tuned using either a TPS or an EO phase modulator. The interferometer is closed by a third 2×2 MMI, which routes the outputs to two edge couplers for off-chip optical collection.
The EO bandwidth of this second-stage interferometer is limited to approximately \SI{1}{\giga\hertz}. 
This limitation arises primarily from packaging-related constraints, namely the lateral pins and long wire bonds, which feature a relatively low RF bandwidth as they are intended for driving components such as TPSs, since the package provides a high-speed RF connection to only one modulator.
Secondly, the electrodes in the MZI are non-terminated. Both of these effects limit the overall bandwidth and lead to the peaking at $\sim$\SI{500}{\mega\hertz} in Fig.~\ref{fig2}c.
While a push-pull configuration was not adopted in this iteration to keep the shorter arm as compact as possible (hundreds of microns), extending the interferometer arms to accommodate such an architecture—while maintaining the necessary path-length imbalance—would improve by a factor of 2 the modulation efficiency ($V_\pi L$). 
We highlight that these limitations are mainly due to packaging constraints, but do not compromise the overall functionalities of the device. Adopting similar design solutions as used for the high-speed optical switch, a comparable performance for the unbalanced MZI could be achieved.

The devices used for the experimental demonstration exhibited an average insertion loss of approximately \SI{6.5}{\dB} due to both coupling and on-chip losses.

The balanced MZM is the key element of the device as it acts as a high-speed optical switch, deterministically routing early (late) time-bin photons to the longer (shorter) path of the interferometer. This configuration eliminates the need for temporal post-selection, enabling direct projection measurements. 
To achieve this, the switch must apply a $\pi$ phase shift within a \SI{100}{\pico\second} time window, corresponding to a modulation frequency of \SI{5}{\giga\hertz}. Thus, the RF line geometry must be carefully optimized to ensure efficient switching performance. The main design criteria include:

\begin{itemize}
    \item Impedance matching between the RF line and the \SI{50}{\ohm} signal source to maximize power transfer.
    \item Velocity matching between the RF phase velocity and the optical group velocity to enhance EO bandwidth.
    \item Field overlap optimization between the RF and optical modes to minimize the modulation voltage ($V_{\pi}$).
\end{itemize}

\begin{figure}[ht]
\centering
\includegraphics[width=0.8\columnwidth]{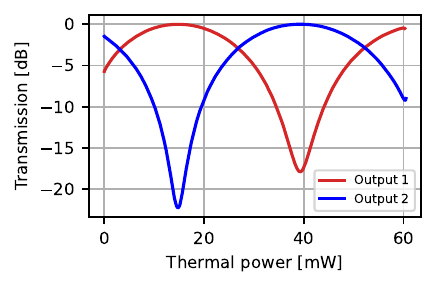}
\caption{
\textbf{ -- Linear characterization.} Transmission spectrum of the unbalanced MZI, obtained by setting the balanced MZM in quadrature to operate as a 50:50 beam splitter, while sweeping the phase of the unbalanced MZI.
}\label{fig_lin_charact}
\end{figure}

At low frequency (\SI{1}{\mega\hertz}), the device exhibits $V_{\pi} = \SI{3.37}{\volt}$, an estimated $V_{\pi} = \SI{4.35}{\volt}$ at \SI{5}{\giga\hertz} and a tested \SI{3}{\decibel} EO bandwidth of approximately \SI{30}{\giga\hertz}.
The ultimate switching frequency of the MZM is estimated to be $\sim \SI{60}{\giga\hertz}$ when modulated with sinusoidal signal for switching, and is primarily limited by the maximum allowable power dissipation in the integrated 50 $\Omega$ RF termination.
Fig.~\ref{fig_lin_charact} shows a linear device characterization, displaying the interference pattern of the unbalanced MZI. In this case, when a continuous-wave (CW) laser is injected, the first balanced MZM is set in quadrature, and the output powers of the two outputs are recorded by sweeping the TPSs, registering an extinction ratio of approximately \SI{20}{\decibel}. 
The relatively limited extinction ratio of the second-stage interferometer, which bounds the maximum achievable quantum interference visibility, is mainly attributed to polarization-mode mixing and differential losses between the two interferometer arms, rather than to any splitting imbalance of the 2×2 MMI (see Supplementary Note 5).
The average thermal power required to achieve a $\pi$ phase shift using the TPSs is $P_{\pi} =$ \SI{23.5}{\milli\watt} among the realized devices.

To ensure robustness and portability, the TFLN chip is fully integrated into a metal package, as shown in Fig.~\ref{fig1}e. 
The optical interface consists of a pigtailed 4-channel fiber array unit (FAU) with Nufern UHNA7 fibers spliced to standard single-mode and polarization maintaining fibers. 
The chip is mounted in the metal package using thermally conductive glue, improving heat dissipation and thermal stabilization. A standard G3PO RF connector is used to feed high-speed modulation signals to the MZM electrodes via precision wire bonding. 
All TPSs and the unbalanced MZI EO phase modulator are connected to external drivers through lateral package pins, which are wire-bonded to the corresponding chip contact pads.

\begin{figure*}[ht]
\centering
\includegraphics[width=1\textwidth]{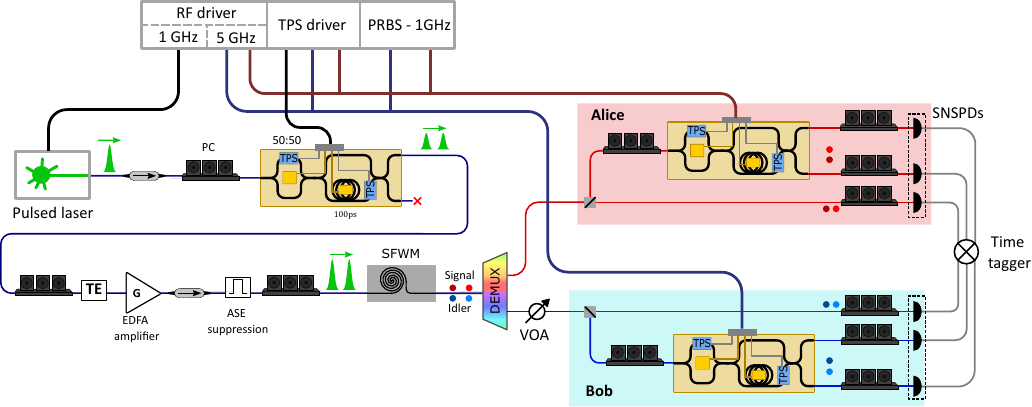}
\caption{
\textbf{ -- Experimental setup schematic. }
PC: polarization controller, TPS: thermal phase shifter, TE: transverse electric polarizer filter, ASE: amplified spontaneous emission, EDFA: erbium-doped fiber amplifier, SFWM: spontaneous four-wave mixing, VOA: variable optical attenuator, SNSPD: superconductive nanowire single photon detector, PRBS: Pseudorandom Binary Sequence, DEMUX: Optical demultiplexer. 
}\label{fig_setup}
\end{figure*}

\vspace{1ex}
\noindent {\bf Experimental setup.} 
The experimental setup used for the experimental demonstration is shown in Fig.~\ref{fig_setup}. For the generation of entangled states, it includes an actively mode locked laser operating at a \SI{1}{\giga\hertz} repetition rate, with a spectral bandwidth of \SI{0.42}{\nano\meter} FWHM corresponding to \SI{9.2}{\pico\second} pulse duration centered at ITU channel 33 (\SI{1550.92}{\nano\meter}). 
The laser pulses pass through an optical isolator, and the polarization is adjusted to match the transverse electric (TE) mode of the fabricated TFLN photonic circuit.
These pulses are then injected into the TFLN device, where the MZM is operated as a 50:50 beam splitter by tuning the TPS, generating two balanced time-bin pump pulses.

After exiting the device, the twin pulses are amplified using an erbium-doped fiber amplifier (EDFA), which also allows tunability of the pump power. 
The ASE noise from both the laser and the amplifier is effectively suppressed by several bandpass filters centered at the laser frequency.
The filtered twin pulses are coupled via grating couplers into a \SI{16}{\milli\meter} long silicon waveguide, realized on \SI{310}{\nano\meter} SOI layer thickness by the CEA-Leti pilot line, where pairs of entangled photons are generated by SFWM.
At the output, the photon pairs are spectrally separated with two bandpass filters centered at ITU channels 28 and 38 (1554.94 nm and 1546.92 nm, respectively). This separates the signal and idler photons, which are then forwarded to Alice and Bob's receivers.

The receiver configuration changes slightly depending on the performed application. In all configurations, polarization controllers are used before and after the device to match the TE polarization and maximize SNSPDs efficiency. 
All three TFLN devices (one for pulse tailoring and two at the receivers) are connected to an external power supply to control the bias. For the balanced MZM, power is applied to only one of the two integrated TPSs while the second is grounded, as a single phase shifter provides a sufficient tuning range for quadrature biasing. The TPS of the unbalanced MZI is controlled in a similar manner. 
We emphasize that no active feedback control system was required for any of the experimental measurements to compensate for bias drift. The device packaging, combined with temperature stabilization via a TEC system, is sufficiently stable against phase drifts. This approach simplifies the operating architecture by eliminating the need for complex auxiliary bias-tracking systems.
The two receiver devices are also connected to an RF signal generator, which applies a sinusoidal voltage at \SI{5}{\giga\hertz} with \SI{4.35}{\volt} peak-to-peak amplitude, enabling operation of the optical switch. 
The RF source also drives the pulsed laser, ensuring synchronization among the two receiver users and the entangled photons source.
The output signals from the SNSPDs are analyzed using a time tagger that records the detection events. The described setup is used for entanglement certification and quantum state tomography.

In QKD with passive basis selection, a fiber based 50:50 beam splitter is placed before the device in order to perform measurements in the computational basis $Z$. To perform the QKD implementation while varying the channel link losses, a VOA is placed before Bob’s beam splitter, which has \SI{2}{\decibel} insertion loss at minimum attenuation.
In the QKD with active basis selection, the beam splitter is removed, and the EO modulator in the unbalanced MZI is driven by an AWG (synchronized with the RF signal generator) to actively control the projective measurements.

\vspace{1ex}

\section*{\small Acknowledgements}
A.B. acknowledges PNRR MUR project F13C22000920005.
M.B. and M.C. acknowledge support of the Italian Ministry of Education (MUR) PNRR project PE0000023-NQSTI. 
M.G. acknowleges the PNRR MUR project CN00000013-HPC.
S.C. acknowledges European Union funding from the STARLight project (project ID: 101194170).
D.B. acknowledges the HyperSpace project (project ID 101070168).
M.C. acknowledges funding from the European Union under the MSCA Postdoctoral Fellowship grant 101211100 (project GLINT).
M.R.B. acknowledges support from the European Union’s Horizon Europe Framework Programme under the Marie Sklodowska Curie Grant No. 101072637, Project Quantum-Safe Internet (QSI).
The authors acknowledge CEA-Leti, Grenoble for providing the silicon photonic chip used as source of photon pairs.

\section*{\small Data availability}
Data underlying the results presented in this paper are not publicly available at this time but may be obtained from the authors upon reasonable request.

\section*{\small Competing interests}
{The authors declare no competing interests.}

\section*{\small Author contributions}
D.B. and M.G. conceived the conceptual idea.
A.B., F.G., A.M., M.P., A.W. and F.A.S. engineered and fabricated the device.
S.C. realized and optimized the Silicon source. 
A.B. and M.B. performed the experimental measurements, under the coordination of M.C., M.G. and D.B.
A.B. performed the data analysis, with support from M.R.B., C.A, G.V., and P.V.
A.B. and M.C. wrote the manuscript, with support from M.R.B.
All authors have proofread the manuscript.

\section*{\protect\small Additional information}
{\small {\bf Supplementary information} Supplementary information accompanies the manuscript.}

\begingroup
\scriptsize	
\bibstyle{naturemag} 
\bibliography{References}

\begin{thebibliography}{10}
\expandafter\ifx\csname url\endcsname\relax
  \def\url#1{\texttt{#1}}\fi
\expandafter\ifx\csname urlprefix\endcsname\relax\def\urlprefix{Available at }\fi
\providecommand{\bibinfo}[2]{#2}
\providecommand{\eprint}[2][]{\url{#2}}
\makeatletter
\providecommand \doibase [0]{https://doi.org/}
\providecommand \@sanitize@url [0]{\catcode `\\12\catcode `\$12\catcode
  `\&12\catcode `\#12\catcode `\^12\catcode `\_12\catcode `\%12\relax}
\providecommand \Doi [0]{\begingroup \@sanitize@url \@Doi }%
\providecommand \@Doi [1]{\endgroup\@@startlink{\doibase#1}\@@Doi}%
\def\gobbledotblock , {}
\let\@@Doi\relax
\newcommand \@@Doi [1]{\textcolor{blue}{\gobbledotblock#1}\@@endlink}
\newcommand\myhrefnoop[2]{\gobbledotblock#2}
\newcommand\@@myhref[2]{\href{#1}{\gobbledotblock#2}}
\newcommand\@myhref[1]{\endgroup\@@myhref{#1}}
\newcommand\myhref{\begingroup\@sanitize@url\@myhref}
\providecommand\@@startlink[1]{%
\leavevmode%
\pdfstartlink\pdfstartlink@attr%
 user{/Subtype/Link/A<</Type/Action/S/URI/URI(#1)>>}\relax%
}
\providecommand\@@endlink{\pdfendlink}
\providecommand\pdfstartlink@attr{attr{/Border[0 0 0 [1 5] ]/H/I/C[0 1 1]}}
\makeatother


\bibitem{Schrdinger1936}
\bibinfo{author}{Schr\"{o}dinger, E.}
\newblock \bibinfo{title}{Probability relations between separated systems}.
\newblock \Doi {10.1017/s0305004100019137} {, \emph{\bibinfo{journal}{Mathematical Proceedings of the Cambridge Philosophical Society}} \textbf{\bibinfo{volume}{32}}, \bibinfo{pages}{446–452} (\bibinfo{year}{1936})}.

\bibitem{Nielsen2012}
\bibinfo{author}{Nielsen, M.~A.} \& \bibinfo{author}{Chuang, I.~L.}
\newblock \emph{\bibinfo{title}{Quantum Computation and Quantum Information}} (\bibinfo{publisher}{Cambridge: Cambridge University Press}, \bibinfo{year}{2010}).

\bibitem{Pittaluga2025}
\bibinfo{author}{Pittaluga, M.} \emph{et~al.}
\newblock \bibinfo{title}{Long-distance coherent quantum communications in deployed telecom networks}.
\newblock \Doi {10.1038/s41586-025-08801-w} {, \emph{\bibinfo{journal}{Nature}} \textbf{\bibinfo{volume}{640}}, \bibinfo{pages}{911–917} (\bibinfo{year}{2025})}.

\bibitem{ekert92}
\bibinfo{author}{Ekert, A.~K.} \emph{et~al.}
\newblock \bibinfo{title}{Practical quantum cryptography based on two-photon interferometry}.
\newblock \Doi {10.1103/PhysRevLett.69.1293} {, \emph{\bibinfo{journal}{Physical Review Letters}} \textbf{\bibinfo{volume}{69}}, \bibinfo{pages}{1293--1295} (\bibinfo{year}{1992})}.

\bibitem{bennett92}
\bibinfo{author}{Bennett, C.~H.}, \bibinfo{author}{Brassard, G.} \& \bibinfo{author}{Mermin, N.~D.}
\newblock \bibinfo{title}{Quantum cryptography without bell's theorem}.
\newblock \Doi {10.1103/PhysRevLett.68.557} {, \emph{\bibinfo{journal}{Physical Review Letters}} \textbf{\bibinfo{volume}{68}}, \bibinfo{pages}{557--559} (\bibinfo{year}{1992})}.

\bibitem{bennett2014}
\bibinfo{author}{Bennett, C.~H.} \& \bibinfo{author}{Brassard, G.}
\newblock \bibinfo{title}{Quantum cryptography: Public key distribution and coin tossing}.
\newblock \Doi {https://doi.org/10.1016/j.tcs.2014.05.025} {, \emph{\bibinfo{journal}{Theoretical Computer Science}} \textbf{\bibinfo{volume}{560}}, \bibinfo{pages}{7--11} (\bibinfo{year}{2014})}.

\bibitem{lo2014secure}
\bibinfo{author}{Lo, H.~K.}, \bibinfo{author}{Curty, M.} \& \bibinfo{author}{Tamaki, K.}
\newblock \bibinfo{title}{Secure quantum key distribution}.
\newblock \Doi {10.1038/nphoton.2014.149} {, \emph{\bibinfo{journal}{Nature Photonics}} \textbf{\bibinfo{volume}{8}}, \bibinfo{pages}{595--604} (\bibinfo{year}{2014})}.

\bibitem{ma2007quantum}
\bibinfo{author}{Ma, X.~F.}, \bibinfo{author}{Fung, C. H.~F.} \& \bibinfo{author}{Lo, H.~K.}
\newblock \bibinfo{title}{Quantum key distribution with entangled photon sources}.
\newblock \Doi {10.1103/physreva.76.012307} {, \emph{\bibinfo{journal}{Physical Review A}} \textbf{\bibinfo{volume}{76}}, \bibinfo{pages}{012307} (\bibinfo{year}{2007})}.

\bibitem{Franson1989}
\bibinfo{author}{Franson, J.~D.}
\newblock \bibinfo{title}{Bell inequality for position and time}.
\newblock \Doi {10.1103/physrevlett.62.2205} {, \emph{\bibinfo{journal}{Physical Review Letters}} \textbf{\bibinfo{volume}{62}}, \bibinfo{pages}{2205–2208} (\bibinfo{year}{1989})}.

\bibitem{tittel1998violation}
\bibinfo{author}{Tittel, W.} \emph{et~al.}
\newblock \bibinfo{title}{Violation of bell inequalities by photons more than 10 km apart}.
\newblock \Doi {10.1103/PhysRevLett.81.3563} {, \emph{\bibinfo{journal}{Physical review letters}} \textbf{\bibinfo{volume}{81}}, \bibinfo{pages}{3563--3566} (\bibinfo{year}{1998})}.

\bibitem{Brendel1999}
\bibinfo{author}{Brendel, J.} \emph{et~al.}
\newblock \bibinfo{title}{Pulsed energy-time entangled twin-photon source for quantum communication}.
\newblock \Doi {10.1103/physrevlett.82.2594} {, \emph{\bibinfo{journal}{Physical Review Letters}} \textbf{\bibinfo{volume}{82}}, \bibinfo{pages}{2594–2597} (\bibinfo{year}{1999})}.

\bibitem{Marcikic2002}
\bibinfo{author}{Marcikic, I.} \emph{et~al.}
\newblock \bibinfo{title}{Time-bin entangled qubits for quantum communication created by femtosecond pulses}.
\newblock \Doi {10.1103/PhysRevA.66.062308} {, \emph{\bibinfo{journal}{Physical Review A}} \textbf{\bibinfo{volume}{66}}, \bibinfo{pages}{062308} (\bibinfo{year}{2002})}.

\bibitem{montaut2025progress}
\bibinfo{author}{Montaut, N.} \emph{et~al.}
\newblock \bibinfo{title}{Progress in integrated and fiber optics for time-bin based quantum information processing}.
\newblock \Doi {10.3389/aot.2025.1560084} {, \emph{\bibinfo{journal}{Advanced Optical Technologies}} \textbf{\bibinfo{volume}{14}}, \bibinfo{pages}{1560084} (\bibinfo{year}{2025})}.

\bibitem{Singh2025}
\bibinfo{author}{Singh, A.} \emph{et~al.}
\newblock \bibinfo{title}{Photonic quantum information with time-bins: principles and applications}.
\newblock \myhrefnoop {} {, \emph{\bibinfo{journal}{Print at https://arxiv.org/abs/2507.08102}}  (\bibinfo{year}{2025})}.

\bibitem{yu2025quantum}
\bibinfo{author}{Yu, H.} \emph{et~al.}
\newblock \bibinfo{title}{Quantum key distribution implemented with d-level time-bin entangled photons}.
\newblock \Doi {10.1038/s41467-024-55345-0} {, \emph{\bibinfo{journal}{Nature Communications}} \textbf{\bibinfo{volume}{16}}, \bibinfo{pages}{171} (\bibinfo{year}{2025})}.

\bibitem{chapman2022hyperentangled}
\bibinfo{author}{Chapman, J.~C.}, \bibinfo{author}{Lim, C. C.~W.} \& \bibinfo{author}{Kwiat, P.~G.}
\newblock \bibinfo{title}{Hyperentangled time-bin and polarization quantum key distribution}.
\newblock \myhrefnoop {} {, \emph{\bibinfo{journal}{Physical Review Applied}} \textbf{\bibinfo{volume}{18}}, \bibinfo{pages}{044027} (\bibinfo{year}{2022})}.

\bibitem{congia2025generation}
\bibinfo{author}{Congia, S.} \emph{et~al.}
\newblock \bibinfo{title}{Generation of hyperentangled photon pairs in the time and frequency domain on a silicon photonic chip}.
\newblock \Doi {10.1364/ol.562079} {, \emph{\bibinfo{journal}{Optics Letters}} \textbf{\bibinfo{volume}{50}}, \bibinfo{pages}{5117} (\bibinfo{year}{2025})}.

\bibitem{Patel2012}
\bibinfo{author}{Patel, K.~A.} \emph{et~al.}
\newblock \bibinfo{title}{Coexistence of high-bit-rate quantum key distribution and data on optical fiber}.
\newblock \Doi {10.1103/PhysRevX.2.041010} {, \emph{\bibinfo{journal}{Physical Review X}} \textbf{\bibinfo{volume}{2}}, \bibinfo{pages}{041010} (\bibinfo{year}{2012})}.

\bibitem{zhang2008generation}
\bibinfo{author}{Zhang, Q.} \emph{et~al.}
\newblock \bibinfo{title}{Generation of 10-ghz clock sequential time-bin entanglement}.
\newblock \Doi {10.1364/OE.16.003293} {, \emph{\bibinfo{journal}{Optics express}} \textbf{\bibinfo{volume}{16}}, \bibinfo{pages}{3293--3298} (\bibinfo{year}{2008})}.

\bibitem{Mueller:24}
\bibinfo{author}{Mueller, A.} \emph{et~al.}
\newblock \bibinfo{title}{High-rate multiplexed entanglement source based on time-bin qubits for advanced quantum networks}.
\newblock \Doi {10.1364/OPTICAQ.509335} {, \emph{\bibinfo{journal}{Optica Quantum}} \textbf{\bibinfo{volume}{2}}, \bibinfo{pages}{64--71} (\bibinfo{year}{2024})}.

\bibitem{Fitzke2022}
\bibinfo{author}{Fitzke, E.} \emph{et~al.}
\newblock \bibinfo{title}{Scalable network for simultaneous pairwise quantum key distribution via entanglement-based time-bin coding}.
\newblock \Doi {10.1103/PRXQuantum.3.020341} {, \emph{\bibinfo{journal}{PRX Quantum}} \textbf{\bibinfo{volume}{3}}, \bibinfo{pages}{020341} (\bibinfo{year}{2022})}.

\bibitem{Kim2022}
\bibinfo{author}{Kim, J.~H.} \emph{et~al.}
\newblock \bibinfo{title}{Quantum communication with time-bin entanglement over a wavelength-multiplexed fiber network}.
\newblock \Doi {10.1063/5.0073040} {, \emph{\bibinfo{journal}{APL Photonics}} \textbf{\bibinfo{volume}{7}}, \bibinfo{pages}{016106} (\bibinfo{year}{2022})}.

\bibitem{Huang2025}
\bibinfo{author}{Huang, Y.~W.} \emph{et~al.}
\newblock \bibinfo{title}{A sixteen-user time-bin entangled quantum communication network with fully connected topology}.
\newblock \Doi {10.1002/lpor.202301026} {, \emph{\bibinfo{journal}{Laser \& Photonics Reviews}} \textbf{\bibinfo{volume}{19}}, \bibinfo{pages}{2301026} (\bibinfo{year}{2025})}.

\bibitem{finco2024time}
\bibinfo{author}{Finco, G.} \emph{et~al.}
\newblock \bibinfo{title}{Time-bin entangled bell state generation and tomography on thin-film lithium niobate}.
\newblock \Doi {10.1038/s41534-024-00925-7} {, \emph{\bibinfo{journal}{npj Quantum Information}} \textbf{\bibinfo{volume}{10}}, \bibinfo{pages}{135} (\bibinfo{year}{2024})}.

\bibitem{maeder2026programmable}
\bibinfo{author}{Maeder, A.} \emph{et~al.}
\newblock \bibinfo{title}{Programmable bell state generation in an integrated thin film lithium niobate circuit}.
\newblock \myhrefnoop {} {, \emph{\bibinfo{journal}{Light: Science \& Applications}} \textbf{\bibinfo{volume}{15}}, \bibinfo{pages}{43} (\bibinfo{year}{2026})}.

\bibitem{Aerts1999}
\bibinfo{author}{Aerts, S.} \emph{et~al.}
\newblock \bibinfo{title}{Two-photon franson-type experiments and local realism}.
\newblock \Doi {10.1103/PhysRevLett.83.2872} {, \emph{\bibinfo{journal}{Physical Review Letters}} \textbf{\bibinfo{volume}{83}}, \bibinfo{pages}{2872--2875} (\bibinfo{year}{1999})}.

\bibitem{jogenfors2015hacking}
\bibinfo{author}{Jogenfors, J.} \emph{et~al.}
\newblock \bibinfo{title}{Hacking the bell test using classical light in energy-time entanglement--based quantum key distribution}.
\newblock \Doi {10.1126/sciadv.1500793} {, \emph{\bibinfo{journal}{Science Advances}} \textbf{\bibinfo{volume}{1}}, \bibinfo{pages}{e1500793} (\bibinfo{year}{2015})}.

\bibitem{xavier2025energy}
\bibinfo{author}{Xavier, G.~B.} \emph{et~al.}
\newblock \bibinfo{title}{Energy-time and time-bin entanglement: past, present and future}.
\newblock \myhrefnoop {} {, \emph{\bibinfo{journal}{npj Quantum Information}} \textbf{\bibinfo{volume}{11}}, \bibinfo{pages}{129} (\bibinfo{year}{2025})}.

\bibitem{Vedovato2018}
\bibinfo{author}{Vedovato, F.} \emph{et~al.}
\newblock \bibinfo{title}{Postselection-loophole-free bell violation with genuine time-bin entanglement}.
\newblock \Doi {10.1103/PhysRevLett.121.190401} {, \emph{\bibinfo{journal}{Physical Review Letters}} \textbf{\bibinfo{volume}{121}}, \bibinfo{pages}{190401} (\bibinfo{year}{2018})}.

\bibitem{Verstraete2002}
\bibinfo{author}{Verstraete, F.} \& \bibinfo{author}{Wolf, M.~M.}
\newblock \bibinfo{title}{Entanglement versus bell violations and their behavior under local filtering operations}.
\newblock \Doi {10.1103/PhysRevLett.89.170401} {, \emph{\bibinfo{journal}{Physical Review Letters}} \textbf{\bibinfo{volume}{89}}, \bibinfo{pages}{170401} (\bibinfo{year}{2002})}.

\bibitem{clauser1969proposed}
\bibinfo{author}{Clauser, J.~F.} \emph{et~al.}
\newblock \bibinfo{title}{Proposed experiment to test local hidden-variable theories}.
\newblock \myhrefnoop {} {, \emph{\bibinfo{journal}{Physical review letters}} \textbf{\bibinfo{volume}{23}}, \bibinfo{pages}{880--884} (\bibinfo{year}{1969})}.

\bibitem{takesue2007entanglement}
\bibinfo{author}{Takesue, H.} \emph{et~al.}
\newblock \bibinfo{title}{Entanglement generation using silicon wire waveguide}.
\newblock \Doi {10.1063/1.2814040} {, \emph{\bibinfo{journal}{Applied Physics Letters}} \textbf{\bibinfo{volume}{91}}, \bibinfo{pages}{201108} (\bibinfo{year}{2007})}.

\bibitem{altepeter2005photonic}
\bibinfo{author}{Altepeter, J.~B.}, \bibinfo{author}{Jeffrey, E.~R.} \& \bibinfo{author}{Kwiat, P.~G.}
\newblock \bibinfo{title}{Photonic state tomography}.
\newblock \Doi {10.1016/S1049-250X(05)52003-2} {, \emph{\bibinfo{journal}{Advances in atomic, molecular, and optical physics}} \textbf{\bibinfo{volume}{52}}, \bibinfo{pages}{105--159} (\bibinfo{year}{2005})}.

\bibitem{Kuzucu2008}
\bibinfo{author}{Kuzucu, O.} \emph{et~al.}
\newblock \bibinfo{title}{Joint temporal density measurements for two-photon state characterization}.
\newblock \Doi {10.1103/physrevlett.101.153602} {, \emph{\bibinfo{journal}{Physical Review Letters}} \textbf{\bibinfo{volume}{101}}, \bibinfo{pages}{153602} (\bibinfo{year}{2008})}.

\bibitem{Drago2024}
\bibinfo{author}{Drago, C.} \& \bibinfo{author}{Sipe, J.~E.}
\newblock \bibinfo{title}{Deconstructing squeezed light: Schmidt decomposition versus the whittaker-shannon interpolation}.
\newblock \Doi {10.1103/PhysRevA.110.023710} {, \emph{\bibinfo{journal}{Physical Review A}} \textbf{\bibinfo{volume}{110}}, \bibinfo{pages}{023710} (\bibinfo{year}{2024})}.

\bibitem{Borghi2024}
\bibinfo{author}{Borghi, M.} \emph{et~al.}
\newblock \bibinfo{title}{Uncorrelated photon pair generation from an integrated silicon nitride resonator measured by time-resolved coincidence detection}.
\newblock \Doi {10.1364/ol.527965} {, \emph{\bibinfo{journal}{Optics Letters}} \textbf{\bibinfo{volume}{49}}, \bibinfo{pages}{3966} (\bibinfo{year}{2024})}.

\bibitem{wootters1998entanglement}
\bibinfo{author}{Wootters, W.~K.}
\newblock \bibinfo{title}{Entanglement of formation of an arbitrary state of two qubits}.
\newblock \myhrefnoop {} {, \emph{\bibinfo{journal}{Physical Review Letters}} \textbf{\bibinfo{volume}{80}}, \bibinfo{pages}{2245--2248} (\bibinfo{year}{1998})}.

\bibitem{RevModPhys.74.145}
\bibinfo{author}{Gisin, N.} \emph{et~al.}
\newblock \bibinfo{title}{Quantum cryptography}.
\newblock \Doi {10.1103/RevModPhys.74.145} {, \emph{\bibinfo{journal}{Reviews of Modern Physics}} \textbf{\bibinfo{volume}{74}}, \bibinfo{pages}{145--195} (\bibinfo{year}{2002})}.

\bibitem{lo2005efficient}
\bibinfo{author}{Lo, H.-K.}, \bibinfo{author}{Chau, H.~F.} \& \bibinfo{author}{Ardehali, M.}
\newblock \bibinfo{title}{Efficient quantum key distribution scheme and a proof of its unconditional security}.
\newblock \Doi {10.1007/s00145-004-0142-y} {, \emph{\bibinfo{journal}{Journal of Cryptology}} \textbf{\bibinfo{volume}{18}}, \bibinfo{pages}{133--165} (\bibinfo{year}{2005})}.

\bibitem{tomamichel2012tight}
\bibinfo{author}{Tomamichel, M.} \emph{et~al.}
\newblock \bibinfo{title}{Tight finite-key analysis for quantum cryptography}.
\newblock \Doi {10.1038/ncomms1631} {, \emph{\bibinfo{journal}{Nature communications}} \textbf{\bibinfo{volume}{3}}, \bibinfo{pages}{634} (\bibinfo{year}{2012})}.

\bibitem{mannalath2025sharp}
\bibinfo{author}{Mannalath, V.}, \bibinfo{author}{Zapatero, V.} \& \bibinfo{author}{Curty, M.}
\newblock \bibinfo{title}{Sharp finite statistics for quantum key distribution}.
\newblock \Doi {10.1103/l735-x48g} {, \emph{\bibinfo{journal}{Physical Review Letters}} \textbf{\bibinfo{volume}{135}}, \bibinfo{pages}{020803} (\bibinfo{year}{2025})}.

\bibitem{pelet2025entanglement}
\bibinfo{author}{Pelet, Y.} \emph{et~al.}
\newblock \bibinfo{title}{Entanglement-based clock syntonization for quantum key distribution networks: Demonstration over a 50 km-long link}.
\newblock \myhrefnoop {} {, \emph{\bibinfo{journal}{Applied Physics Letters}} \textbf{\bibinfo{volume}{126}}, \bibinfo{pages}{174003} (\bibinfo{year}{2025})}.

\bibitem{williams2021implementation}
\bibinfo{author}{Williams, J.} \emph{et~al.}
\newblock \bibinfo{title}{Implementation of quantum key distribution and quantum clock synchronization via time bin encoding}.
\newblock \myhrefnoop {} {, \emph{\bibinfo{journal}{{Proceedings of SPIE 11699, Quantum Computing, Communication, and Simulation}}} }\bibinfo{note}{SPIE, 2021, 16--25}.

\bibitem{calderaro2020fast}
\bibinfo{author}{Calderaro, L.} \emph{et~al.}
\newblock \bibinfo{title}{Fast and simple qubit-based synchronization for quantum key distribution}.
\newblock \myhrefnoop {} {, \emph{\bibinfo{journal}{Physical Review Applied}} \textbf{\bibinfo{volume}{13}}, \bibinfo{pages}{054041} (\bibinfo{year}{2020})}.

\bibitem{acin2007device}
\bibinfo{author}{Ac{\'\i}n, A.} \emph{et~al.}
\newblock \bibinfo{title}{Device-independent security of quantum cryptography against collective attacks}.
\newblock \myhrefnoop {} {, \emph{\bibinfo{journal}{Physical Review Letters}} \textbf{\bibinfo{volume}{98}}, \bibinfo{pages}{230501} (\bibinfo{year}{2007})}.

\bibitem{peri2026high}
\bibinfo{author}{Peri, A.} \emph{et~al.}
\newblock \bibinfo{title}{High-performance heterodyne receiver for quantum information processing in a laser-written integrated photonic platform}.
\newblock \myhrefnoop {} {, \emph{\bibinfo{journal}{Advanced Photonics}} \textbf{\bibinfo{volume}{8}}, \bibinfo{pages}{016009--016009} (\bibinfo{year}{2026})}.

\bibitem{bertapelle2025high}
\bibinfo{author}{Bertapelle, T.} \emph{et~al.}
\newblock \bibinfo{title}{High-speed source-device-independent quantum random number generator on a chip}.
\newblock \myhrefnoop {} {, \emph{\bibinfo{journal}{Optica Quantum}} \textbf{\bibinfo{volume}{3}}, \bibinfo{pages}{111--118} (\bibinfo{year}{2025})}.

\bibitem{Santagiustina2024}
\bibinfo{author}{Santagiustina, F. B.~L.} \emph{et~al.}
\newblock \bibinfo{title}{Experimental post-selection loophole-free time-bin and energy-time nonlocality with integrated photonics}.
\newblock \Doi {10.1364/OPTICA.499247} {, \emph{\bibinfo{journal}{Optica}} \textbf{\bibinfo{volume}{11}}, \bibinfo{pages}{498--511} (\bibinfo{year}{2024})}.

\bibitem{marcikic2004distribution}
\bibinfo{author}{Marcikic, I.} \emph{et~al.}
\newblock \bibinfo{title}{Distribution of time-bin entangled qubits over 50 km of optical fiber}.
\newblock \myhrefnoop {} {, \emph{\bibinfo{journal}{Physical review letters}} \textbf{\bibinfo{volume}{93}}, \bibinfo{pages}{180502} (\bibinfo{year}{2004})}.

\bibitem{Honjo:08}
\bibinfo{author}{Honjo, T.} \emph{et~al.}
\newblock \bibinfo{title}{Long-distance entanglement-based quantum key distribution over optical fiber}.
\newblock \Doi {10.1364/OE.16.019118} {, \emph{\bibinfo{journal}{Optics Express}} \textbf{\bibinfo{volume}{16}}, \bibinfo{pages}{19118--19126} (\bibinfo{year}{2008})}.

\bibitem{pelet2023operational}
\bibinfo{author}{Pelet, Y.} \emph{et~al.}
\newblock \bibinfo{title}{Operational entanglement-based quantum key distribution over 50 km of field-deployed optical fibers}.
\newblock \myhrefnoop {} {, \emph{\bibinfo{journal}{Physical Review Applied}} \textbf{\bibinfo{volume}{20}}, \bibinfo{pages}{044006} (\bibinfo{year}{2023})}.

\bibitem{de2005long}
\bibinfo{author}{de~Riedmatten, H.} \emph{et~al.}
\newblock \bibinfo{title}{Long-distance entanglement swapping with photons from separated sources}.
\newblock \myhrefnoop {} {, \emph{\bibinfo{journal}{Physical Review A}} \textbf{\bibinfo{volume}{71}}, \bibinfo{pages}{050302} (\bibinfo{year}{2005})}.

\bibitem{tagliavacche2025frequency}
\bibinfo{author}{Tagliavacche, N.} \emph{et~al.}
\newblock \bibinfo{title}{Frequency-bin entanglement-based quantum key distribution}.
\newblock \myhrefnoop {} {, \emph{\bibinfo{journal}{npj Quantum Information}} \textbf{\bibinfo{volume}{11}}, \bibinfo{pages}{60} (\bibinfo{year}{2025})}.

\bibitem{zhang2024compact}
\bibinfo{author}{Zhang, J.} \emph{et~al.}
\newblock \bibinfo{title}{Compact polarization splitter-rotator based on lithium niobate-on-insulator platform}.
\newblock \myhrefnoop {} {, \emph{\bibinfo{journal}{Journal of Applied Physics}} \textbf{\bibinfo{volume}{136}}, \bibinfo{pages}{164503} (\bibinfo{year}{2024})}.

\bibitem{sheng2022one}
\bibinfo{author}{Sheng, Y.-B.}, \bibinfo{author}{Zhou, L.} \& \bibinfo{author}{Long, G.-L.}
\newblock \bibinfo{title}{One-step quantum secure direct communication}.
\newblock \Doi {https://doi.org/10.1016/j.scib.2021.11.002} {, \emph{\bibinfo{journal}{Science Bulletin}} \textbf{\bibinfo{volume}{67}}, \bibinfo{pages}{367--374} (\bibinfo{year}{2022})}.

\bibitem{yang2025300}
\bibinfo{author}{Yang, Y.~L.} \emph{et~al.}
\newblock \bibinfo{title}{A 300-km fully-connected quantum secure direct communication network}.
\newblock \Doi {https://doi.org/10.1016/j.scib.2025.02.038} {, \emph{\bibinfo{journal}{Science Bulletin}} \textbf{\bibinfo{volume}{70}}, \bibinfo{pages}{1445--1451} (\bibinfo{year}{2025})}.

\bibitem{sheng2010deterministic}
\bibinfo{author}{Sheng, Y.-B.} \& \bibinfo{author}{Deng, F.-G.}
\newblock \bibinfo{title}{Deterministic entanglement purification and complete nonlocal bell-state analysis with hyperentanglement}.
\newblock \myhrefnoop {} {, \emph{\bibinfo{journal}{Physical Review A}} \textbf{\bibinfo{volume}{81}}, \bibinfo{pages}{032307} (\bibinfo{year}{2010})}.

\bibitem{Korzh2020}
\bibinfo{author}{Korzh, B.} \emph{et~al.}
\newblock \bibinfo{title}{Demonstration of sub-3 ps temporal resolution with a superconducting nanowire single-photon detector}.
\newblock \Doi {10.1038/s41566-020-0589-x} {, \emph{\bibinfo{journal}{Nature Photonics}} \textbf{\bibinfo{volume}{14}}, \bibinfo{pages}{250–255} (\bibinfo{year}{2020})}.

\bibitem{jiang2018low}
\bibinfo{author}{Jiang, X.~H.}, \bibinfo{author}{Wu, H.} \& \bibinfo{author}{Dai, D.~X.}
\newblock \bibinfo{title}{Low-loss and low-crosstalk multimode waveguide bend on silicon}.
\newblock \myhrefnoop {} {, \emph{\bibinfo{journal}{Optics express}} \textbf{\bibinfo{volume}{26}}, \bibinfo{pages}{17680--17689} (\bibinfo{year}{2018})}.

\bibitem{wang2024lithium}
\bibinfo{author}{Wang, C.~L.} \emph{et~al.}
\newblock \bibinfo{title}{Lithium tantalate photonic integrated circuits for volume manufacturing}.
\newblock \myhrefnoop {} {, \emph{\bibinfo{journal}{Nature}} \textbf{\bibinfo{volume}{629}}, \bibinfo{pages}{784--790} (\bibinfo{year}{2024})}.

\bibitem{ekert1991quantum}
\bibinfo{author}{Ekert, A.~K.}
\newblock \bibinfo{title}{Quantum cryptography based on bell’s theorem}.
\newblock \myhrefnoop {} {, \emph{\bibinfo{journal}{Physical review letters}} \textbf{\bibinfo{volume}{67}}, \bibinfo{pages}{661--663} (\bibinfo{year}{1991})}.

\bibitem{zhuang2025ultrabright}
\bibinfo{author}{Zhuang, S.-C.} \emph{et~al.}
\newblock \bibinfo{title}{Ultrabright entanglement based quantum key distribution over a 404 km optical fiber}.
\newblock \myhrefnoop {} {, \emph{\bibinfo{journal}{Physical Review Letters}} \textbf{\bibinfo{volume}{134}}, \bibinfo{pages}{230801} (\bibinfo{year}{2025})}.

\bibitem{lim2021security}
\bibinfo{author}{Lim, C. C.~W.} \emph{et~al.}
\newblock \bibinfo{title}{Security analysis of quantum key distribution with small block length and its application to quantum space communications}.
\newblock \myhrefnoop {} {, \emph{\bibinfo{journal}{Physical Review Letters}} \textbf{\bibinfo{volume}{126}}, \bibinfo{pages}{100501} (\bibinfo{year}{2021})}.

\bibitem{zhong2024hyperentanglement}
\bibinfo{author}{Zhong, Z.-Q.} \emph{et~al.}
\newblock \bibinfo{title}{Hyperentanglement quantum communication over a 50 km noisy fiber channel}.
\newblock \myhrefnoop {} {, \emph{\bibinfo{journal}{Optica}} \textbf{\bibinfo{volume}{11}}, \bibinfo{pages}{1056--1061} (\bibinfo{year}{2024})}.

\bibitem{doda2021quantum}
\bibinfo{author}{Doda, M.} \emph{et~al.}
\newblock \bibinfo{title}{Quantum key distribution overcoming extreme noise: Simultaneous subspace coding using high-dimensional entanglement}.
\newblock \myhrefnoop {} {, \emph{\bibinfo{journal}{Physical Review Applied}} \textbf{\bibinfo{volume}{15}}, \bibinfo{pages}{034003} (\bibinfo{year}{2021})}.

\bibitem{neumann2022continuous}
\bibinfo{author}{Neumann, S.~P.} \emph{et~al.}
\newblock \bibinfo{title}{Continuous entanglement distribution over a transnational 248 km fiber link}.
\newblock \myhrefnoop {} {, \emph{\bibinfo{journal}{Nature Communications}} \textbf{\bibinfo{volume}{13}}, \bibinfo{pages}{6134} (\bibinfo{year}{2022})}.

\end{thebibliography}
\endgroup



\section*{Supplementary material}

# Supplementary information for: Gigahertz-rate thin-film lithium niobate receiver for time-bin quantum communication

Andrea Bernardi[1,2,*], Marco Clementi[3], Marcello Bacchi[3], Matías Rubén Bolaños[4], Sara Congia[3,5], Francesco Garrisi[2], Andrea Martellosio[2], Marco Passoni[2], Alexander Wrobel[2], Costantino Agnesi[4,6], Giuseppe Vallone[4,6], Paolo Villoresi[4], Federico Andrea Sabattoli[2], Matteo Galli[3], and Daniele Bajoni[1]

[1]Dipartimento di Ingegneria Industriale e dell'Informazione, Università di Pavia, Via A. Ferrata 5, 27100 Pavia, Italy
[2]Advanced Fiber Resources Milan s.r.l., Via Fellini 4, 20097 San Donato Milanese, Italy
[3]Dipartimento di Fisica "A. Volta", Università di Pavia, Via A. Bassi 6, 27100 Pavia, Italy
[4]Dipartimento di Ingegneria dell'Informazione, Università degli Studi di Padova, Via Gradenigo 6B, 35131 Padua, Italy
[5]CEA LETI and University Grenoble Alpes, MINATEC Campus, F-38054 Grenoble Cedex, France
[6]Quantum Technologies Research Center, Università degli Studi di Padova, Via Gradenigo 6B, 35131 Padua, Italy
[*]Corresponding author. Email: andrea.bernardi01@universitadipavia.it

## Supplementary Note 1: Secure key rate analysis

Both QKD with active and passive basis selection schemes implement the BBM92 protocol. In both cases, the two users perform measurements in two mutually unbiased bases, namely the $X$ and $Z$ bases for the passive case, and the $X$ and $Y$ for the active case. We perform the SKR computation for both the asymptotic regime and for finite block length for a more realistic scenario. We denote by $C_{|ab\rangle}$ the number of detected coincidence events between the two users when Alice and Bob project their measurements onto the states on the basis $B$ with eigenstates $|a\rangle$ and $|b\rangle$, respectively, where $B : a,b \in \{Z : 0,1; X : +,-; Y : R,L\}$. We denote $p_B$ as the probability that Bob and Alice measure the states in the basis $B$. The measurement rates for the three mutually unbiased bases are:

$$R_X = (C_{|++\rangle} + C_{|+-\rangle} + C_{|-+\rangle} + C_{|--\rangle})/\tau,$$
$$R_Y = (C_{|RR\rangle} + C_{|RL\rangle} + C_{|LR\rangle} + C_{|LL\rangle})/\tau,$$
$$R_Z = (C_{|00\rangle} + C_{|01\rangle} + C_{|10\rangle} + C_{|11\rangle})/\tau$$

where $\tau$ is the time required to accumulate a block of key. The corresponding quantum bit error rates (QBER) are given by:

$$Q_Z = \frac{C_{|01\rangle} + C_{|10\rangle}}{C_{|00\rangle} + C_{|11\rangle} + C_{|01\rangle} + C_{|10\rangle}},$$
$$Q_X = \frac{C_{|+-\rangle} + C_{|-+\rangle}}{C_{|++\rangle} + C_{|--\rangle} + C_{|+-\rangle} + C_{|-+\rangle}},$$
$$Q_Y = \frac{C_{|RL\rangle} + C_{|LR\rangle}}{C_{|RR\rangle} + C_{|LL\rangle} + C_{|RL\rangle} + C_{|LR\rangle}}.$$

To obtain the SKR, in the passive basis choice experiment we take advantage of the intrinsic asymmetry of the receiver, with $p_Z > p_X$, to define the $Z$ basis as the one used to generate secret key (so called *key basis*), and the $X$ basis as the one used for parameter estimation (so called *check basis*). On the other hand, for the active basis choice experiment, since both measuring probabilities are equal, i.e. $p_X = p_Y = 0.5$, by construction, it is not straightforward to define one of the two basis as the key/check basis. Thus, we opted to divide the detected coincidences of each basis choice into two groups: 90% of the data in basis $B$ to generate key in said basis, while the remaining 10% is used for parameter estimation on basis $B'$, with $B \neq B'$. The

relative SKR is defined as follows for both passive and active basis choice configurations:

$$S_{pass} = S_Z \tag{S1}$$
$$S_{act} = S_X + S_Y \tag{S2}$$

where, in the asymptotic regime, each basis measurement contribution is given by[1]:

$$S_Z = R_Z[1 - H(Q_X) - fH(Q_Z)] \tag{S3}$$
$$S_Y = R_Y[1 - H(Q_X) - fH(Q_Y)] \tag{S4}$$
$$S_X = R_X[1 - H(Q_Y) - fH(Q_X)] \tag{S5}$$

where $H(p) = -p\log_2 p - (1-p)\log_2(1-p)$ is the binary Shannon entropy function used for the privacy amplification, and $f = 1.06$ accounts for error correction.

In the finite block size regime, we redefine the contribution of each basis to the SKR as[1,2]:

$$S_Z = R_Z[1 - H(Q'_{X,pass}) - fH(Q_Z)] - \log\frac{2}{\varepsilon_{cor}\varepsilon_{sec}^2} \tag{S6}$$

$$S_Y = R_Y[1 - H(Q'_{X,act}) - fH(Q_Y)] - \log\frac{2}{\varepsilon_{cor}\varepsilon_{sec}^2} \tag{S7}$$

$$S_X = R_X[1 - H(Q'_Y) - fH(Q_X)] - \log\frac{2}{\varepsilon_{cor}\varepsilon_{sec}^2} \tag{S8}$$

where $\varepsilon_{cor}, \varepsilon_{sec}$ are the correctness and secrecy parameters respectively. Here, we assumed that $\varepsilon_{cor} = 10^{-10}$ and $\varepsilon_{sec} = 10^{-10}$. (it is important to note that the source-independent nature of EBQKD protocols renders the inefficiencies in the source irrelevant for the secret key analysis). The parameter $Q'_B$ represents the adjusted QBER for basis $B$, accounting for the statistical fluctuations that arise due to the finite block size. In particular, $Q'_{X,pass}(Q'_{X,act})$ corresponds to the adjusted QBER when using $X$ as the check basis for the passive (active) basis choice experiment. These upper bounds when using the Serfling inequality are given by:

$$Q'_{X,pass} = Q_X + \sqrt{\frac{(n_Z+1)\log(1/\varepsilon_{sec})}{2n_X(n_X+n_Z)}} \tag{S9}$$

$$Q'_{X,act} = Q_X + \sqrt{\frac{(n_Y+1)\log(1/\varepsilon_{sec})}{2n_X(n_X+n_Y)}} \tag{S10}$$

$$Q'_Y = Q_Y + \sqrt{\frac{(n_X+1)\log(1/\varepsilon_{sec})}{2n_Y(n_Y+n_X)}} \tag{S11}$$

where $n_B$ accounts for the total coincidences in the basis $B$. Thus they are defined as:

$$n_X = C_{|++\rangle} + C_{|--\rangle} + C_{|+-\rangle} + C_{|-+\rangle},$$
$$n_Y = C_{|RR\rangle} + C_{|LL\rangle} + C_{|RL\rangle} + C_{|LR\rangle},$$
$$n_Z = C_{|00\rangle} + C_{|11\rangle} + C_{|01\rangle} + C_{|10\rangle}$$

Analogously, when using the Chernoff inequality, the adjusted QBER is defined as

$$Q'_{X,pass} = \frac{(n_Z+n_X)\Gamma^+_{n_X,\varepsilon_{PE}}(m_X/n_X) - m_X}{n_Z}, \tag{S12}$$

$$Q'_{X,act} = \frac{(n_Y+n_X)\Gamma^+_{n_X,\varepsilon_{PE}}(m_X/n_X) - m_X}{n_Y}, \tag{S13}$$

$$Q'_Y = \frac{(n_Y+n_X)\Gamma^+_{n_Y,\varepsilon_{PE}}(m_Y/n_Y) - m_Y}{n_X}, \tag{S14}$$

with $\varepsilon_{PE}$ the failure probability of the parameter estimation step, and

$$\Gamma^+_{n,\varepsilon}(p) = \begin{cases} \gamma^+_{n,\varepsilon}(p) & \text{if } p \in \left[0, \frac{1-2\kappa_{n,\varepsilon}}{1+\kappa_{n,\varepsilon}}\right] \\ 1+\varepsilon & \text{otherwise,} \end{cases} \tag{S15}$$

where $\kappa_{n,\varepsilon} = \frac{2}{9n} \ln(1/\varepsilon)$ and $\gamma^+_{n,\varepsilon}(x)$ defined as

$$\gamma^+_{n,\varepsilon}(x) = \frac{1}{1+4\kappa_{n,\varepsilon}} \left[ 3\kappa_{n,\varepsilon} + (1-2\kappa_{n,\varepsilon})x + 3\sqrt{\kappa_{n,\varepsilon}\left(\kappa_{n,\varepsilon}+x-x^2\right)} \right].$$

## Supplementary Note 2: Analysis optimization of detected events on Z basis

In the context of QKD with passive basis selection, the *Z* measurement bases discriminates whether photons are detected in the early or late time-bin. In our setup, the limited temporal resolution of the SNSPDs (characterized by a timing jitter of approximately 50 ps) leads to a non-negligible overlap between the detection distributions of the two time-bins, as shown in the statistical histogram of detection events in Fig. S1a.

To mitigate the resulting ambiguity in time-bin discrimination, we optimize the analysis of detection events by introducing a discarding window, centered between the early and late peaks with a tunable width. Detection events falling within this window are excluded, thereby reducing the QBER excluding ambiguous detections that could be incorrectly assigned to the wrong time-bin. However, this strategy also lowers the total number of valid detection events, as shown in Fig. S1b.

Fig. S1c illustrates how the SKR is affected by the trade off between reduced QBER and decreased detection rate. For the data presented in the Results section, we used an optimized discarding window width of 30 ps, which yielded the best SKR performance for both Chernoff and Serfling bounds.

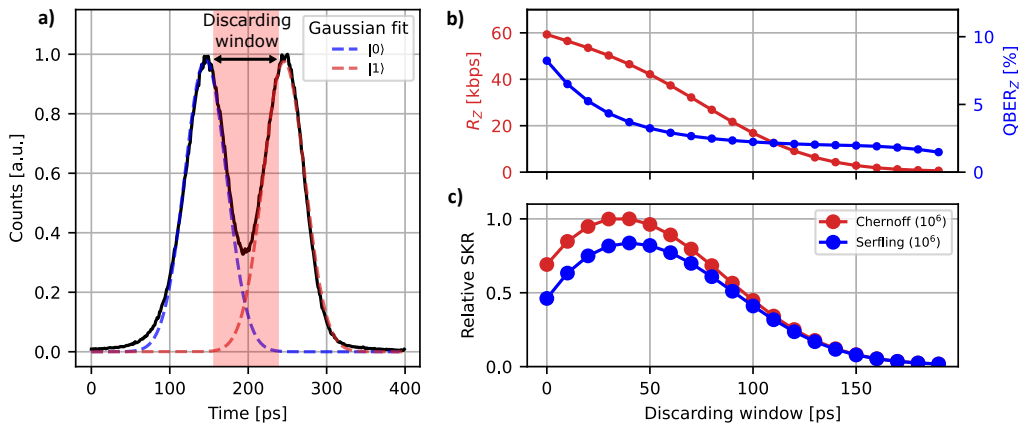

**Fig. S 1 – Measurements on bases Z.** **a** Histogram of time detection events for time-bin encoded photons in the *Z* basis for one QKD user. The detection time is relative to the laser pulse emission. **b** Coincidence rate and QBER between Alice and Bob of time-bin entangled photons in the *Z* basis, as a function of the discarding window width. **c** SKR as a function of the discarding window in the passive basis selection configuration. The optimal value is 30 ps for both bounds with a block size of $10^6$.

## Supplementary Note 3: QKD on variable fiber length

The QKD experimental implementation was also tested by varying the fiber channel length using spools of SMF28 optical fiber. The fiber spool was placed instead of the VOA, thus affecting only the Bob user's channel. This configuration was tested for the passive basis selection scheme. In standard single-mode fiber, the pulses which have a FWHM of 9.2 ps and a spectral bandwidth of 0.42 nm are subject to chromatic dispersion, which causes temporal broadening of the time-bin wavepacket as it propagates. In our case this effect will prevent to discriminate correctly in the basis *Z* the 100 ps separated time-bins, and in case of measurements in the interferometric bases *X* and *Y* to apply the correct switching inside the receiver device.

Fig. S2 shows the histogram detection events as a function of arrival time for both users in the *Z* basis. The temporal shape of Alice's detection events remains unchanged since its channel length is fixed and serves as a reference. In contrast, Bob's detection events broaden with increasing fiber length due to dispersion. This temporal broadening leads to greater overlap between the early and late time-bins, resulting in an increased QBER. To mitigate the effects of chromatic dispersion, and thus enable the practical deployment of the QKD system over long fiber links, a dispersion compensation strategy should be envisioned, for example by the use of dispersion compensating fiber or chirped fiber-Bragg gratings. We emphasize that these strategies, common to standard telecommunication systems, do not introduce, in principle, any disturbance to the quantum channel, except for a typically mild insertion loss.

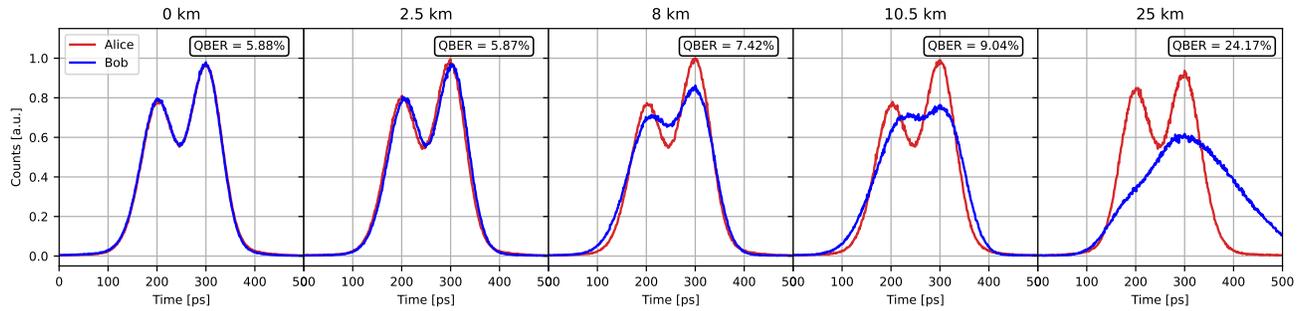

**Fig. S 2 – Dispersion effect on variable fiber length.** Normalized detection counts in the $Z$ basis for the two QKD users, measured at different fiber-link lengths for Bob. The time reference on the x-axis is relative to the pump pulse emission. As the fiber length increases, the pulses broaden, reducing the temporal distinguishability of Bob's time-bin states.

## Supplementary Note 4: Source characterization and pump power optimization

An integrated silicon waveguide fabricated by CEA Leti was used to generate entangled photon pairs. Measuring 16 mm in length, it featured a rib geometry and access via grating couplers. The total insertion loss, including input/output coupling and propagation loss at the pump wavelength of 1550.92 nm, is 5.5 dB. Fig. S3a shows the characterization of the entangled photon generation by means of SFWM, with signal and idler photons at ITU channels 28 and 38, respectively, as a function of the pump pulse energy. The pulse energy inside the waveguide is estimated from the input pump power, taking into account equal input/output coupling losses and the pulse repetition rate.

The retrieved generation rate in waveguide shows the expected quadratic dependence on the pulse energy. The measurement range is limited by the maximum allowed average coupled pump power ($< 15\,\mathrm{dBm}$), which prevented reaching the saturation regime. The Coincidence to accidental (CAR) ratio was computed as $CAR = (C_C - C_A)/C_A$ where $C_C$ and $C_A$ are relatively the detection events associated to the entangled photons, and the accidental detection events associated to coincidences delayed by multiples repetition rate. The integration window conditions were equals for both the detection contributes.

For QKD applications, the pump power should be selected to maximize the final SKR[3]. While a higher mean photon number increases the coincidence rate, it also leads to an increase in the QBER. As shown in Fig. S3b, the SKR for the active basis selection QKD scheme is plotted as a function of the internal source generation probability. The maximum SKR is achieved at an internal pair generation probability of approximately 0.7%. The results reported for the active-basis selection configuration are obtained using a mean photon number of approximately 0.2%, leading to the possibility of further SKR improvements by optimizing the generation rate.

In the case of the passive QKD scheme, a similar optimization can be applied. However, in our experiment we were limited by the maximum detection rate of the SNSPDs ($\sim 1.5\,\mathrm{MHz}$) for measurements in the $Z$ basis.

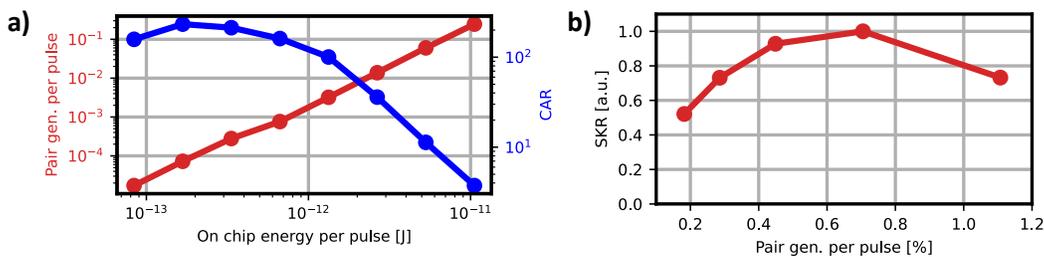

**Fig. S 3 – Source characterization.** **a** Internal waveguide pair-generation probability in ITU channels 28 and 38 (red), and coincidence-to-accidental ratio (CAR) (blue), as a function of the pump pulse energy. **b** Normalized SKR of the QKD protocol with active basis selection as a function of the internal waveguide pair-generation probability per pulse.

## Supplementary Note 5: Additional characterization of the integrated photonic platform

We present additional test measurements of the TFLN technology used in this work to provide further insight into the performance of the integrated optical circuit. In particular, we report statistical measurements of components fabricated on the

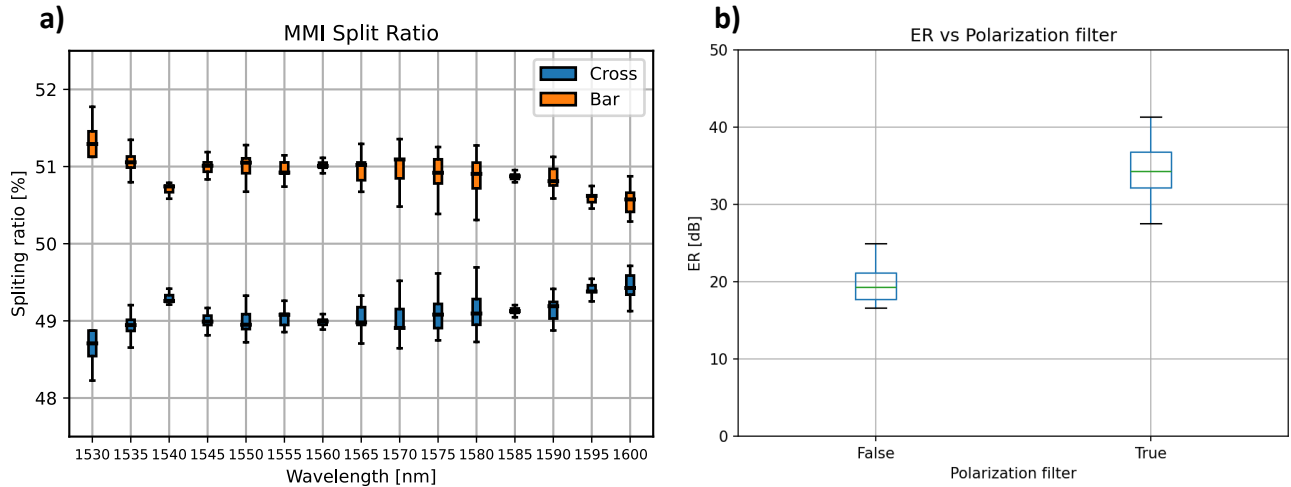

**Fig. S 4 – TFLN components characterization.** **a** Statistical characterization of the power splitting balance of standalone 2×2 MMIs fabricated on the same wafer as the presented device and featuring identical geometry. **b** Statistical distribution of the measured extinction ratio (ER) for interferometric structures fabricated on the same wafer and featuring the same optical components as the tested device, realized with and without an integrated polarization-splitter rotator (PSR). In both panels box rectangles extend from the 25th to 75th percentile, middle line shows the median and external lines show the farthest data point lying within 1.5x the inter-quartile range from the box.

same wafer as the device under test. These include standalone 2×2 MMI components and interferometric devices with an optical design similar to that of the realized device. Two sets of interferometric devices were fabricated to allow for a performance comparison based on the presence or absence of an on-chip polarization-splitter rotator (PSR) utilized as a polarization filter. These test measurements enable the identification and exclusion of factors that may limit interferometric visibility, which in turn affects the achievable quantum interference and the resulting QBER in the two interferometric bases $X$ and $Y$.

The results obtained from the tested MMI components are shown in Fig. S4a. In the tested wavelength range from 1530 nm to 1600 nm, the MMIs exhibit a flat power imbalance of approximately 1%, corresponding to a theoretical extinction ratio (ER) of at least 34 dB. This level of imbalance alone, therefore, does not represent the dominant limitation to the visibility observed in our experiments.

Fig. S4b reports the ER obtained from interferometric structures featuring the same optical components as the present device, realized in two configurations: with and without an integrated PSR acting as a polarization filter to suppress the unwanted TM mode. The data reveal a clear improvement in the achieved ER of more than 10 dB when the PSR is present, providing evidence that polarization-mode mixing is a primary contributor to visibility degradation.

In conclusion, the main factors limiting the quantum interferometric visibility in the present device are the unbalanced losses intrinsic to the second-stage interferometer and polarization-mode mixing. In our implementation, the differential propagation loss is estimated to be approximately $1.3\,\text{cm} \times 0.2\,\text{dB}\,\text{cm}^{-1}$ in the longer arm and about 2% ($\sim 0.1\,\text{dB}$) in the shorter arm, leading to a small amplitude asymmetry between the interferometer paths. Residual TE–TM coupling further contributes to visibility degradation, as evidenced by the improved ER observed in devices incorporating an on-chip polarization filter.

# References

1. Yin, J. *et al.* Entanglement-based secure quantum cryptography over 1,120 kilometres. *Nature* **582**, 501–505 (2020).
2. Tomamichel, M., Lim, C. C. W., Gisin, N. & Renner, R. Tight finite-key analysis for quantum cryptography. *Nature communications* **3**, 634 (2012).
3. Ma, X., Fung, C.-H. F. & Lo, H.-K. Quantum key distribution with entangled photon sources. *Physical Review A—Atomic, Molecular, and Optical Physics* **76**, 012307 (2007).



\end{document}